\documentclass[aps,prd,twocolumn,superscriptaddress,10pt]{revtex4-1}

\usepackage{amsmath}
\usepackage{etoolbox}
\usepackage{graphicx}
\usepackage[usenames,dvipsnames]{xcolor}
\usepackage[colorlinks=true,citecolor=blue,urlcolor=blue]{hyperref}

\newcommand{\g}{\ensuremath{\mathrm{g}}}
\newcommand{\cm}{\ensuremath{\mathrm{cm}}}
\newcommand{\km}{\ensuremath{\mathrm{km}}}
\newcommand{\pc}{\ensuremath{\mathrm{pc}}}
\newcommand{\kpc}{\ensuremath{\mathrm{kpc}}}
\newcommand{\s}{\ensuremath{\mathrm{s}}}
\newcommand{\Gyr}{\ensuremath{\mathrm{Gyr}}}
\newcommand{\Msun}{\ensuremath{\mathrm{M}_{\odot}}}

\begin{document}

\title{Accelerated core collapse in tidally stripped self-interacting dark matter halos}

\author{Hiroya Nishikawa}
\author{Kimberly K. Boddy}
\affiliation{Department of Physics \& Astronomy, Johns Hopkins University, Baltimore, Maryland 21218, USA}
\author{Manoj Kaplinghat}
\affiliation{Department of Physics and Astronomy, University of California, Irvine, California 92697, USA}

\begin{abstract}
We use a semianalytic approach that is calibrated to N-body simulations to study the evolution of self-interacting dark matter cores in galaxies.
We demarcate the regime where the temporal evolution of the core density follows a well-defined track set by the initial halo parameters and the cross section.
Along this track, the central density reaches a minimum value set by the initial halo density.
Further evolution leads to an outward heat transfer, inducing gravothermal core collapse such that the core shrinks as its density increases.
We show that the timescale for the core collapse is highly sensitive to the outer radial density profile.
Satellite galaxies with significant mass loss due to tidal stripping should have larger central densities and significantly faster core collapse compared to isolated halos.
Such a scenario could explain the dense and compact cores of dwarf galaxies in the Local Group like Tucana (isolated from the Milky Way), the classical Milky Way satellite Draco, and some of the ultrafaint satellites.
If the ultimate fate of core collapse is black hole formation, then the accelerated timescale provides a new mechanism for creating intermediate-mass black holes.
\end{abstract}

\maketitle

\section{Introduction}
Self-interacting dark matter (SIDM) \cite{Spergel:1999mh,Firmani:2000ce,Kaplinghat:2015aga} is a compelling framework for explaining the small-scale structure formation puzzles~\cite{Bullock:2017xww}.
SIDM simulations show that elastic collisions of dark matter particles transfer heat towards the colder central regions of dark matter halos, lowering central densities and creating constant density cores~\cite{Vogelsberger:2012ku,Rocha:2012jg,Peter:2012jh,Zavala:2012us,Vogelsberger:2014pda,Fry:2015rta,Fitts:2018ycl}, which has long been recognized as a means of alleviating the longstanding core-cusp problem~\cite{deBlok:2009sp,Salucci:2011ee}.

The core-cusp issue is tricky, because galaxies exhibit a large diversity of rotation curves~\cite{deNaray:2009xj,Oman:2015xda}, while following an approximate radial acceleration relation~\cite{McGaugh:2016leg}.
Recent work has shown that SIDM models with a cross section of few $\cm^2/\g$ at low velocities can solve these interconnected problems for spiral galaxies~\cite{Kamada:2016euw,Ren:2018jpt}.

In this paper, we investigate the evolution of a dark matter halo in the presence of self-interactions using a semianalytic method, originally developed to study gravothermal collapse in globular clusters~\cite{doi:10.1093/mnras/191.3.483,Quinlan:1996bw} and later applied to isolated SIDM halos~\cite{Balberg:2002ue,Koda:2011yb}.
This method allows us to track the full halo evolution at scales $\leq 100~\mathrm{pc}$, which are expensive to achieve with N-body simulations.
We first characterize the full process of SIDM halo evolution, whereby large cores can be created today for cross sections of $\sim\! 1~\cm^2/\g$ and the gravothermal collapse phase sets in for larger cross sections, as seen in simulations.
We show that this temporal evolution is accelerated if the outer region of the halo is stripped, resulting in a higher density core today.

Our key result is that a large cross section leads to central densities that are higher than for field halos if the halo has experienced tidal stripping early in its formation history.
Our results are applicable to local group galaxies that have interacted with the Milky Way (MW) in the past (such as Tucana) and satellites of the MW (including ultrafaint galaxies).
The timescale for halo collapse depends on characteristic density of the initial halo and the details of the truncation process; thus, gravothermal collapse contributes to the diversity of MW satellites, for similar cross sections that can also explain the diversity of spiral galaxies~\cite{Kamada:2016euw,Ren:2018jpt}.
We also discuss the possible consequences of core collapse of the SIDM halo for intermediate-mass black hole (BH) formation.

\begin{figure*}[t]
  \includegraphics[width=\columnwidth]{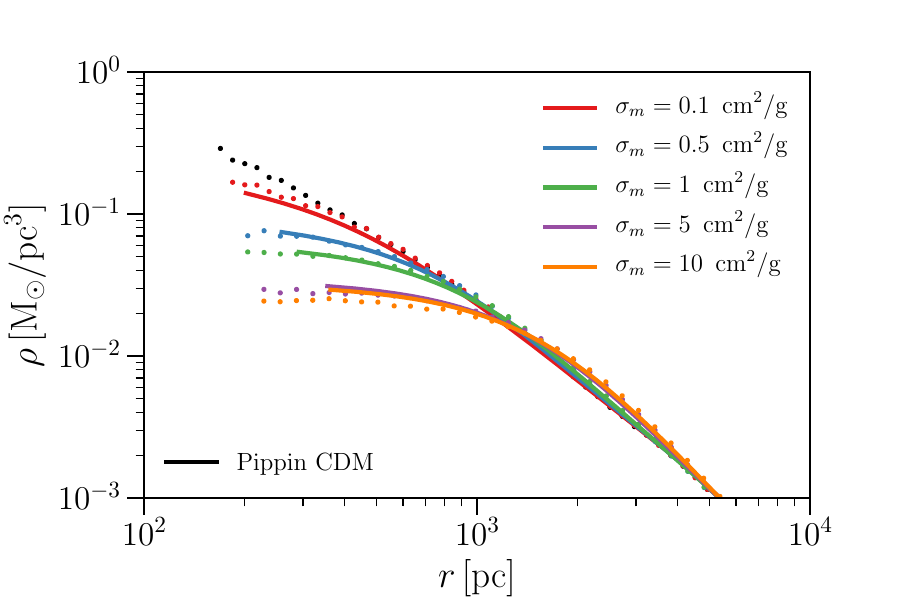}
  \includegraphics[width=\columnwidth]{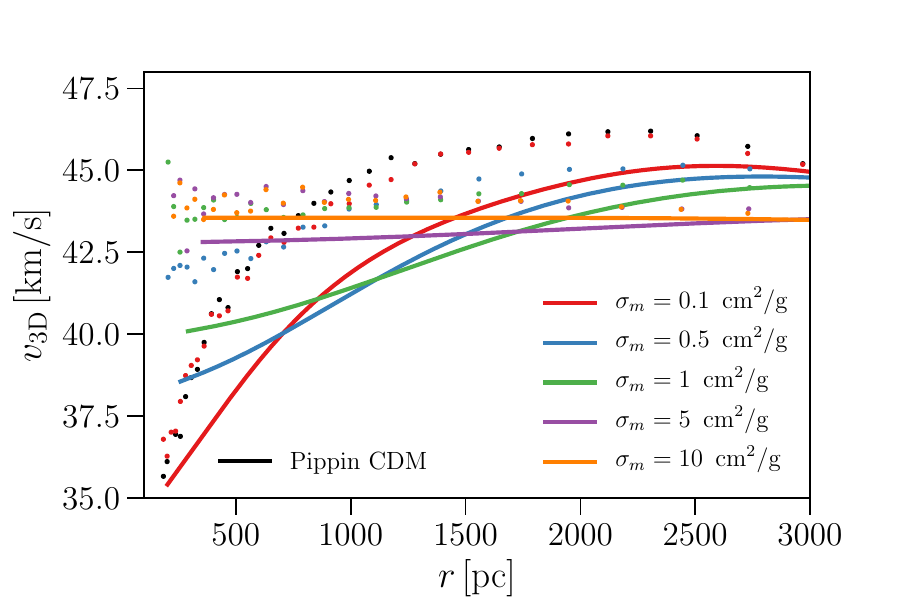}
  \caption{Comparison of the semianalytic approach in this work (solid) with the N-body simulation of the Pippin halo in Ref.~\cite{Elbert:2014bma} (dotted).
    We show the density profile (left panel) and 3D velocity dispersion (right panel) for SIDM cross sections per mass $\sigma_m$ indicated in the legend.
    For the semianalytic calculation, we start with an initial density profile matching that of the Pippin CDM halo (black, dotted) and evolve the halo for $13~\Gyr$.
    We set $C = 0.75$ to best match the properties of the Pippin halos, particularly the density profiles for $\sigma_m= 0.5$ and $1~\cm^2/\g$.}
  \label{fig:Pippin_comparison}
\end{figure*}

\section{Gravothermal fluid model}
We consider a spherical halo with a density profile $\rho(r,t)$ and an enclosed mass of $M(r,t)$ at radius $r$ and time $t$.
We assume that this halo structure is set very early on ($\sim\! \Gyr$), as is relevant for the small mass halos that we focus on in this work.
The halo is assumed to be isotropic and spherically symmetric, and dark matter is modeled as a thermally conducting fluid in quasistatic virial equilibrium.
The dark matter particles with mass $m$ self-interact with a cross section per mass $\sigma_m$.
They have a one-dimensional (1D) velocity dispersion $v(r,t)$ with a corresponding three-dimensional (3D) velocity dispersion $v_{3\textrm{D}}=\sqrt{3}v$.
The relaxation time is defined as $t_r \equiv \lambda/(av)$, where $\lambda = 1/(\rho\sigma_m)$ is the mean-free path and $a=\sqrt{16/\pi}$ is a coefficient relevant for hard-sphere scattering of particles with a Maxwell-Boltzmann velocity distribution~\cite{Balberg:2002ue}.

Through self-interactions, heat can flow from one region of the halo to another with a luminosity $L(r,t)$ through a spherical shell located at radius $r$.
Heat transfer is given by Fourier's law of thermal conduction,
\begin{equation}
  \frac{L}{4\pi r^2} = -\kappa \nabla T \ ,
  \label{eq:conduction}
\end{equation}
where $\kappa$ is the thermal conductivity.
Strictly speaking, this equation is valid in the short mean-free path (SMFP) regime in which $\lambda \ll H$, where $H\equiv\sqrt{v^2/(4\pi G\rho)}$ is the gravitational scale height or Jean's length.
In this case, $\kappa_\textrm{SMFP} = (3/2)b\rho\lambda^2/(a m t_r)$~\cite{Balberg:2002ue}, where $b = 25\sqrt{\pi}/32 \approx 1.38$ is the effective impact parameter, calculated in Chapman-Enskog theory~\cite{chapman1990mathematical}.
In the long mean-free path (LMFP) regime in which $\lambda \gg H$, the thermal conduction formula empirically well-describes the gravothermal collapse of globular clusters with $\kappa_\textrm{LMFP} = (3/2)C\rho H^2 / (m t_r)$~\cite{doi:10.1093/mnras/191.3.483}, where $C$ is a calibration parameter, described below.
We interpolate between these regimes via $\kappa^{-1} = \kappa^{-1}_\textrm{SMFP} + \kappa^{-1}_\textrm{LMFP}$~\cite{Balberg:2002ue}.

Using the above ansatz, we can write the heat conduction equation in dimensionless variables as:
\begin{equation}
  \begin{gathered}
    \frac{\partial\tilde{L}}{\partial\tilde{r}}
    = -\tilde{r}^2 \tilde{\rho} \tilde{v}^2
    \left(\frac{\partial}{\partial\tilde{t}}\right)_{\tilde{M}}
    \ln\left(\frac{\tilde{v}^3}{\tilde{\rho}}\right) \\
    \tilde{L} = -\frac{3}{2} \tilde{r}^2 \tilde{v}
    \left(\frac{a}{b} {\tilde{\sigma}_m}^2 + \frac{1}{C}
     \frac{1}{\tilde{\rho}\tilde{v}^2}\right)^{-1}
    \frac{\partial \tilde{v}^2}{\partial \tilde{r}} \ ,
  \end{gathered}
  \label{eq:heat_transfer}
\end{equation}
where $\tilde{r} \equiv r/r_s$,  $\tilde{v} \equiv v/v_0$, $\tilde{M} \equiv M/M_0$, $\tilde{t} \equiv t/t_0$, $\tilde{\sigma}_m \equiv \sigma_m/(4\pi r_s^2 M_0^{-1})$, $\tilde{\rho}\equiv \rho/\rho_s$, and $\tilde{L}\equiv L/[(G M_0^2){(r_s t_0)}^{-1}]$.
We use the mass scale $M_0 =  4\pi r_s^3 \rho_s$, velocity scale $v_0 = \sqrt{GM_0/r_s}$, and timescale $t_0^{-1}=a \sigma_m v_0 \rho_s$.

\begin{figure*}[t]
  \includegraphics[width=\columnwidth]{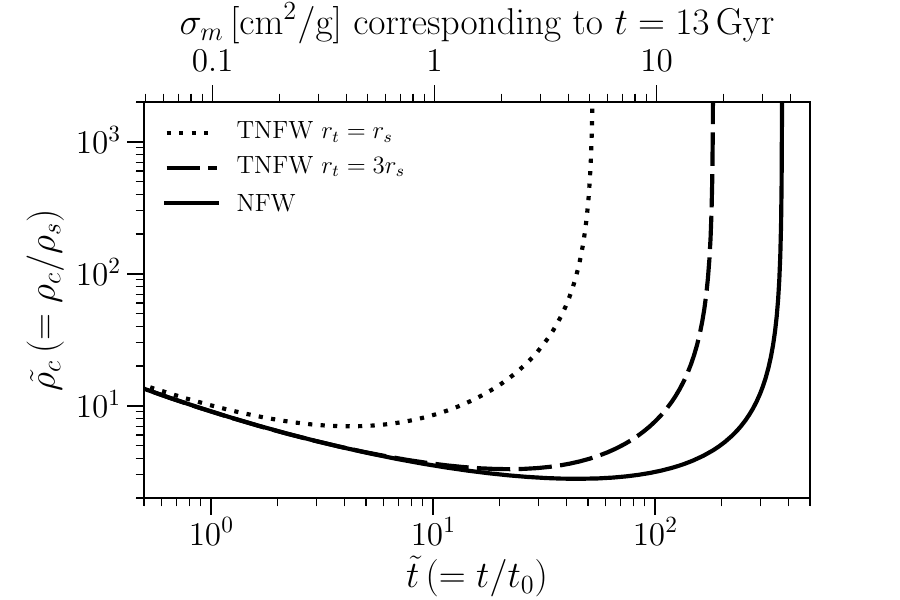}
  \includegraphics[width=\columnwidth]{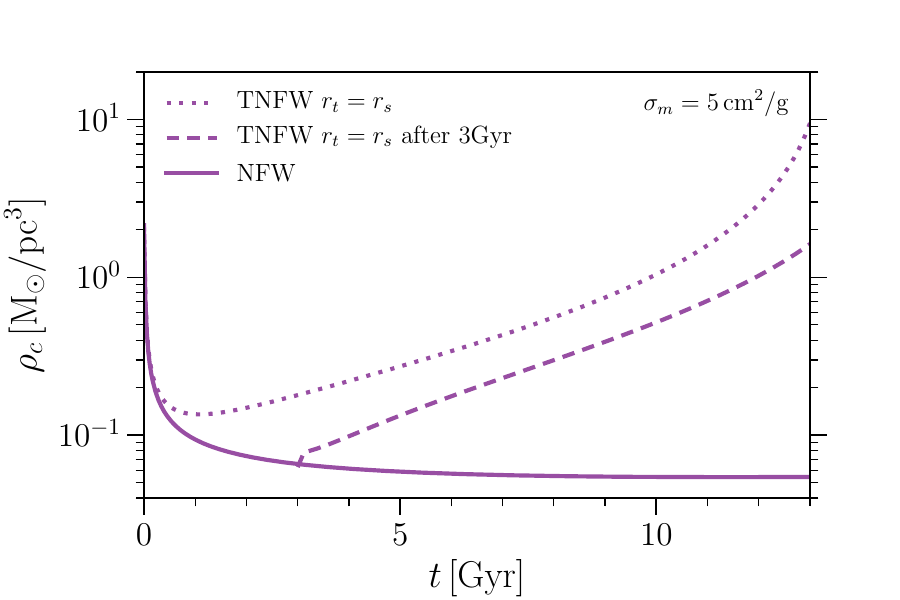}
  \caption{\textbf{[Left]} Central density evolutions for an initially NFW profile (solid) and initially truncated NFW (TNFW) profiles, with truncation radii $r_t=r_s$ (dotted) and $r_t=3\, r_s$ (dashed).
    These curves apply to halos with $\sigma_m \leq 10~\cm^2/\g$, for which the LMFP regime is dominant throughout the majority of the evolution shown here.
    The central densities reach a minimum of $\tilde{\rho} = 7.1$, $3.3$, and $2.8$ at times $\tilde{t} = 4.3$, $22$, and $45$ for the TNFW with $r_t=r_s$, TNFW with $r_t=3\, r_s$, and NFW profiles, respectively.
    Since $t_0$ depends on $\sigma_m$, halos with different values of $\sigma_m$ (shown on the top axis, in units of $\cm^2/\g$) are at different locations along these evolution curves by today (physical time of $13~\Gyr$).
    \textbf{[Right]} Central density evolutions for $\sigma_m = 5~\cm^2/\g$ for an initially NFW profile (solid), an initially TNFW profile with $r_t=r_s$ (dotted), and an initially NFW profile truncated at $r_t=r_s$ after $3~\Gyr$ of evolution (dashed).}
  \label{fig:central_density_evolution}
\end{figure*}

We numerically solve Eq.~\eqref{eq:heat_transfer} along with the equations for mass conservation and hydrostatic equilibrium, using the techniques described in Ref.~\cite{Pollack:2014rja}.
See Appendix~\ref{app:implementation} for details of our numerical implementation.

We calibrate $C$ using the Pippin SIDM halos in the LMFP regime~\cite{Elbert:2014bma}.
We solve the gravothermal fluid equations using the Pippin CDM halo directly [as opposed to a Navarro-Frenk-White (NFW) fit to it] as our initial condition.
As shown in Fig.~\ref{fig:Pippin_comparison}, we find that $C=0.75$ produces halo density profiles that are in reasonable agreement with those from Pippin for $\sigma_m \leq 10~\cm^2/\g$, while the 3D velocity dispersions are systematically lower.
The semianalytic approach enforces hydrostatic equilibrium, which does not hold in the simulation, particularly for small cross sections.
Nonetheless, the velocity dispersions match qualitative features towards the center of the halo.
In particular, for the largest cross sections shown, the halo is isothermal and the velocity dispersion has a negative radial gradient.

We note that the calibration depends on the simulation (e.g., isolated halo~\cite{Koda:2011yb} versus cosmological~\cite{Elbert:2014bma}).
As $C$ controls the importance of the LMFP regime, increasing (decreasing) its value causes a delay (advance) in the time of core collapse.
For the range of $\sigma_m$ under consideration, however, mildly changing the calibration does not strongly affect the agreement in Fig.~\ref{fig:Pippin_comparison}; see Appendix~\ref{app:calibration}.
Additionally, deep in the LMFP regime, a change in $C$ can be compensated by a change in $\sigma_m$.
Thus, while the precise values of $\sigma_m$ in our subsequent analyses are subject to systematic uncertainties of the calibration, our main results comparing halos with and without truncation (as we now discuss) still hold.

\section{Tidally stripped halos}
We proceed to investigate the consequences of self-interactions on tidally stripped halos.
Tidal effects impact both MW satellites (e.g., Tucana III~\cite{Li:2018zfs,Simon:2016mkr} and Triangulum II~\cite{Kirby:2015bxa}) and field dwarfs (e.g., Tucana~\cite{Teyssier:2012ay}).
This process strips away stars and the outer regions of the subhalo and is thus potentially applicable to the dwarf galaxies on sufficiently radial orbits that experience significant mass loss.

For a profile $\rho_{\mathrm{NFW}}(r) = \rho_s/[(r/r_s){(1+(r/r_s))}^2]$, we model the effect of tidal stripping by changing the profile for $r> r_t$ to $\rho_\mathrm{NFW}(r_t) \times (r_t/r)^{p_t}$, where $p_t=5$~\cite{Penarrubia:2010jk}.
This simple model is meant to illustrate the effects of mass loss from the outer region, not to represent a realistic description of SIDM subhalo orbital history.
We consider truncations at $r_t = r_s$ and $3\, r_s$, corresponding to a halo mass loss of $\sim \! 90\%$ and $\sim \! 70\%$, respectively, for a typical concentration of $20$.
A halo truncated at the same $r_t$ with a shallower outer slope of $p_t=4$ undergoes core collapse slightly later, but the effect is qualitatively similar to $p_t=5$.
We choose $\rho_s = 0.019~\Msun/\pc^3$ and $r_s = 2.59~\kpc$, obtained from fitting the Pippin CDM profile to an NFW profile using the \textsc{Colossus} cosmology code~\cite{Diemer:2017bwl}.

\section{Temporal evolution of central density}
We focus on the central density $\rho_c$, defined as the average density of the innermost region of the halo specified in our analysis (at $\tilde{r} < 0.01$, as described in Appendix~\ref{app:implementation}).
We show the evolution of $\rho_c$ for initially NFW and initially TNFW profiles in the left panel of Fig.~\ref{fig:central_density_evolution}.
For the cross sections of interest, the halos remain in the LMFP limit well into the runaway phase of collapse (the nearly vertical portion of the curves).
In the LMFP regime, Eq.~\eqref{eq:heat_transfer} is independent of the value of $\tilde{\sigma}_m$ and thus the gravothermal evolution shown holds for all $\sigma_m \leq 10~\cm^2/\g$.

For all three profiles, the central density drops as the core of the halo forms and rises again as the core begins to collapse.
The minimum core density occurs approximately when the luminosity $L$ in the central region of the halo transitions from being negative (positive temperature gradient) to being positive (negative temperature gradient).
The minimum density for halos with no tidal stripping is about $3 \, \rho_s$, independent of the cross section.
For the cases with tidal stripping, the collapse time becomes shorter and the minimum core density is higher.
For the NFW profile, only for cross sections $\sigma_m \gtrsim 4~\cm^2/\g$ do we find the central density rise as the core begins to collapse within the lifetime of the Universe.
However, for the TNFW profile with truncation at $r_t=r_s$ and $r_t=3\, r_s$, cross sections of $\sigma_m \gtrsim 0.4~\cm^2/\g$ and $2~\cm^2/\g$, respectively, have started to collapse by today.

To roughly gauge the impact of infall time, we allow the halo to evolve as before for a period of time before abruptly truncating it.
We neglect the impact of multiple pericenter passages in our simplified analysis.
We show the central density evolution of a halo truncated at $r_t=r_s$ after $3~\Gyr$ (or $z\simeq 2$) for an SIDM cross section of $\sigma_m = 5~\cm^2/\g$ in the right panel of Fig.~\ref{fig:central_density_evolution}.
Truncation times of $3-6~\Gyr$ are appropriate for the closest MW dwarfs~\cite{Rocha:2011aa}.
Such an extreme tidal stripping event leads to almost 2 orders of magnitude increase in the central density, showing the importance of this effect for nearby satellites.
Note that an initial truncation at $r_t=3\, r_s$ barely alters the central density evolution away from the NFW case within the lifetime of the Universe, as seen from the left panel of Fig.~\ref{fig:central_density_evolution} for $\sigma_m = 5~\cm^2/\g$.

After truncation, hydrostatic equilibrium significantly lowers the pressure of the halo beyond the point of truncation, where most of the mass is lost, causing the velocity dispersion (and thus temperature) to decrease substantially.
Heat flows towards the colder truncated part of the halo from the region near the scale radius, where the temperature is highest.
As a result, heat is diffused more quickly within the truncated halo, leading to a faster formation of the isothermal core and thus an accelerated evolution for core collapse.
We provide a more detailed description of the halo evolution in Appendix~\ref{app:evolution}.

\section{Observational consequences for the local group}
Simulated CDM halos have circular velocities that are systematically higher than those observed for field dwarf galaxies~\cite{Garrison-Kimmel:2014vqa} and MW satellites~\cite{BoylanKolchin:2011dk}, while dark matter self-interactions reduce circular velocities to the observed range of $10-20~\km/\s$~\cite{Elbert:2014bma}.
A notable exception is the field dwarf Tucana~\cite{Fraternali:2009de,Kirby:2014sya}, which has been isolated for a long time ($\sim\! 10~\Gyr$) but is observed to have a high circular velocity $v_{\mathrm{circ}}\simeq33.7~\km/\s$ at $r_{1/2}\simeq 0.28~\kpc$~\cite{Kirby:2014sya,Garrison-Kimmel:2014vqa}, even for CDM~\cite{Fitts:2018ycl,Garrison-Kimmel:2014vqa}.

Accounting for truncation from tidal stripping, we have shown that the central density of a halo is larger than expected from CDM (if it has begun the process of core collapse), with a correspondingly larger circular velocity.
Indeed, the TNFW halo with a delayed truncation shown in the right panel of Fig.~\ref{fig:central_density_evolution} produces a circular velocity of $v_{\mathrm{circ}}=34~\km/\s$ at the same radius, in line with observations of Tucana.
Our suggestion that Tucana may have experienced large tidal effects early on is consistent with a proposal to explain the isolation of Tucana~\cite{Sales:2007hr}, wherein Tucana came in with a companion and was ejected due to three-body interactions, while the companion became bound to the MW (and perhaps fully disrupted).
This early ejection scenario also seems consistent with the kinematics of Tucana~\cite{Fraternali:2009de} and the early cutoff in star formation~\cite{Saviane:1996xf,2010ApJ...722.1864M}.

Similar to Tucana, the MW satellite Draco (dwarf spheroidal) has a large central density~\cite{Read:2018pft}, which may be difficult to explain with cross sections $\sigma_m > 1~\cm^2/\g$, assuming evolution like a field halo~\cite{Valli:2017ktb}.
Draco most likely had a close pericenter passage according to \textit{Gaia} data~\cite{2018A&A...619A.103F} and hence our arguments would suggest it could also have enhanced its central density compared to the field SIDM halo evolution.
Note that the presence of the disk will increase tidal stripping and enhance the effects we have discussed~\cite{DOnghia:2009xhq,Garrison-Kimmel:2017zes,Kelley:2018pdy}.

The ultrafaint satellites of the MW are likely to have even smaller pericenters~\cite{2018ApJ...863...89S}, owing to their large velocities and closer distances.
These galaxies would certainly have been impacted by the shortened timescales and higher core densities predicted by our semianalytic method.
A subhalo with $V_{\max}=10~\km/\s$ that is tidally truncated at $r=r_s$ at around $z=2$ would evolve to have a smaller $r_{\rm max}$ and a denser core by up to an order of magnitude (compared to a similar field SIDM halo) around $30-50~\pc$ radius (typical half-light radii for ultrafaints).
The denser core may have a bearing on the survival of these subhalos~\cite{Penarrubia:2010jk}, and the adiabatic contraction of the stars (as the SIDM density increases) would impact the compactness of the stellar distribution.

\begin{figure}[t]
  \includegraphics[width=\columnwidth]{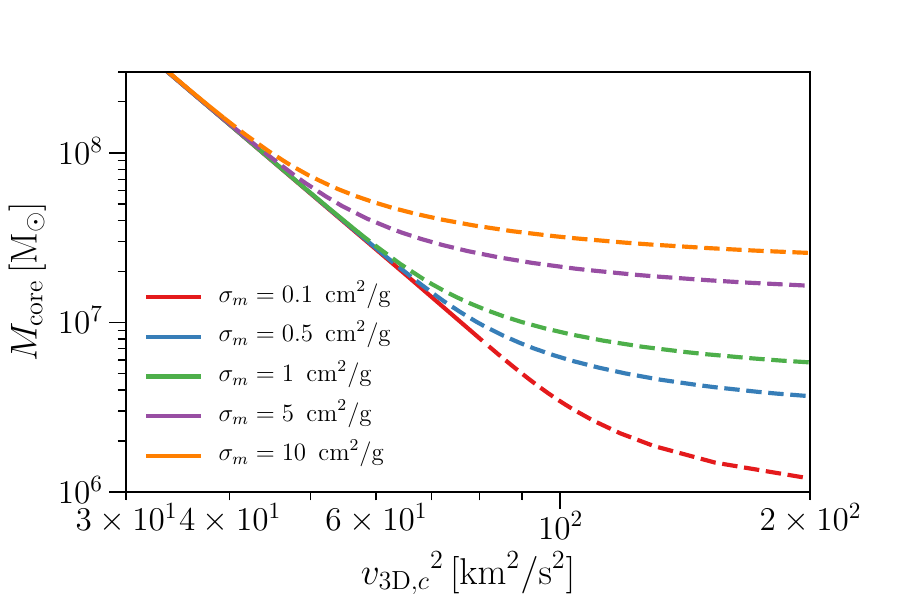}
  \caption{Core mass as a function of the central 3D velocity dispersion for an NFW profile.
    As the dispersion increases, the core sheds its mass rapidly while it remains in LMFP regime (solid).
    Once the core enters SMFP regime ($\kappa_{\mathrm{SMFP}} \leq \kappa_{\mathrm{LMFP}}$), the mass loss decreases (dashed).
    Note that the SMFP phase of the evolution takes place in a very short amount of time compared to the LMFP phase.
    The same process occurs within the TNFW profiles at much earlier times.}
  \label{fig:rho_core_temperature}
\end{figure}

\section{Intermediate-mass black holes}
We note that the mass in the central region of the halo remains unaffected by a tidal stripping, which should only affect the halo at distances near or beyond the scale radius.
Defining the core mass $M_\textrm{core}$ to be the mass within a core radius $r_\textrm{core}$ at which $\rho(r_\textrm{core}) = 0.1\, \rho_c$, we show the relation between $M_\textrm{core}$ and $v_{3\textrm{D},c}$ in Fig.~\ref{fig:rho_core_temperature} after the core has begun to collapse.
This relation should hold for both truncated and field halos.

When the halo is in the LMFP regime, $\log(M_\textrm{core})$ decreases as $\log({v_{3\textrm{D}, c}}^2)$ increases, with a slope of $-4.17$, in agreement with the self-similar collapse scenario~\cite{Balberg:2002ue}.
Eventually, the SMFP heat transfer in the core begins to take over and the core becomes optically thick, inhibiting the diffusion of heat and mass from the core and thus resulting in a lower rate of mass loss.

An interesting consequence of the runaway gravothermal collapse in the SMFP regime is the possible formation of a BH~\cite{Balberg:2001qg,Pollack:2014rja,Choquette:2018lvq}, which may exist today due to accelerated collapse from tidal stripping.
If we define the core mass $M_{\mathrm{SMFP}}$ to be the mass within the radius at which $\kappa_{\textrm{SMFP}}=\kappa_{\textrm{LMFP}}$, then we can apply previous work that found $\mathrm{d} \log (M_{\mathrm{SMFP}}) /\mathrm{d} \log({v_{3\textrm{D}, c}}^2) \simeq -0.85$ asymptotically \cite{Balberg:2001qg,Balberg:2002ue}.
If a BH should form at the onset of relativistic instability at $v_{3\textrm{D}, c} \approx c/3$~\cite{Balberg:2001qg,Balberg:2002ue}, then $M_{\mathrm{SMFP}} = 10^7 - 10^8 \, \Msun$ at $v_{3\textrm{D}, c} = 60~\km/\s$ (see Table~\ref{table:SMFP_core} in Appendix~\ref{app:evolution}) would create BHs with mass $M_{\mathrm{BH}} =  30 - 300 \, \Msun$.
This estimate involves a large extrapolation of our results and should be considered as motivation for further work rather than a concrete prediction.

Depending on the mass accretion history, these black holes may grow further.
Finally, for rare halos that collapse early (with high concomitant densities), it is possible that this rapid gravothermal collapse may result in seeds for supermassive BHs, without assuming $\sigma_m \gg 1$ for a subdominant dark matter fraction~\cite{Pollack:2014rja,Choquette:2018lvq}.

The exact properties of the SIDM cross section (e.g., velocity dependence, dissipative interactions, etc.) may become relevant as the extrapolation spans over orders of magnitude in $v_{c}$, and such interactions could remove particles from the collapsing core~\cite{Chu:2018nki,Vogelsberger:2018bok,Essig:2018pzq,Choquette:2018lvq}.
Consideration of the angular momentum of SIDM particles may also require a closer examination of the final stage of collapse.

\section{Conclusions}
We applied the gravothermal fluid model, calibrated to cosmological SIDM simulations~\cite{Elbert:2014bma}, to tidally stripped halos.
We showed that for $\sigma_m < 10~\cm^2/\g$, the halo core density follows a track defined by the outer halo parameters ($\rho_s$ and $r_s$), attaining a minimum value of $\sim 3 \, \rho_s$.
We found that mass loss from the outer regions can shorten the timescale of this evolution and increase the minimum SIDM density.

The accelerated evolution due to tidal stripping opens up the possibility of intermediate-mass BH formation in the rapid gravothermal collapse phase in dwarf halos for $\sigma_m \gtrsim 5~\cm^2/\g$.
The increase in the core density due to gravothermal collapse could explain the high central density of the isolated Local Group dwarf Tucana, assuming it came close to the MW about $10~\Gyr$ ago.
The same arguments suggest that the central dark matter density of satellites that have come close to the MW disk, like Draco and the ultrafaint dwarfs, would be enhanced for large cross sections.
In SIDM models with a cross section per mass of a few $\cm^2/\g$, which are favored by fits to the rotation curves of field galaxies \cite{Ren:2018jpt}, our results suggest that the structural properties of satellite galaxies should be correlated with their orbital histories.

\begin{acknowledgments}
We thank Anna Kwa and Annika Peter, who contributed in the early stages of this work.
We also thank Jason Pollack for useful conversations.
HN was supported by NSF Grant No.\ 1519353, and MK was supported by NSF Grant No.\ PHY1620638.
\end{acknowledgments}

\appendix
\section{Numerical implementation}
\label{app:implementation}
We describe our procedure, which follows that in Ref.~\cite{Pollack:2014rja}, for solving the gravothermal evolution equations
\begin{subequations}
  \begin{gather}
    \frac{\partial\tilde{M}}{\partial\tilde{r}}
    = \tilde{r}^2 \tilde{\rho} \label{eq:mass_conservation}\\
    \frac{\partial (\tilde{\rho}\tilde{v}^2)}{\partial\tilde{r}}
    = -\frac{\tilde{M} \tilde{\rho}}{\tilde{r}^2} \label{eq:hydrostatic}\\
    \frac{\partial\tilde{L}}{\partial\tilde{r}}
    = -\tilde{r}^2 \tilde{\rho} \tilde{v}^2
    \left(\frac{\partial}{\partial\tilde{t}}\right)_{\tilde{M}}
    \ln\left(\frac{\tilde{v}^3}{\tilde{\rho}}\right) \label{eq:thermo}\\
    \tilde{L}
    = -\frac{3}{2} \tilde{r}^2 \tilde{v}
    \left(\frac{a}{b} {\tilde{\sigma}_m}^2 + \frac{1}{C}
     \frac{1}{\tilde{\rho}\tilde{v}^2}\right)^{-1}
    \frac{\partial \tilde{v}^2}{\partial \tilde{r}} \ , \label{eq:gravothermal}
  \end{gather}
\end{subequations}
which describe mass conservation, hydrostatic equilibrium, the first law of thermodynamics, and heat conduction, respectively, in dimensionless variables.
For an initial density profile of the NFW form, the characteristic scales of the halo are the NFW scale density $\rho_s$ and radius $r_s$.
With these quantities, we write the dimensionless radius $\tilde{r} \equiv r/r_s$, 1D velocity dispersion $\tilde{v} \equiv v/v_0$, cross section per mass $\tilde{\sigma}_m \equiv \sigma_m/(4\pi r_s^2 M_0^{-1})$, density $\tilde{\rho}\equiv \rho/\rho_s$, and luminosity $\tilde{L}\equiv L/[(G M_0^2){(r_s t_0)}^{-1}]$.
We use the mass scale $M_0 =  4\pi r_s^3 \rho_s$, velocity scale $v_0 = \sqrt{GM_0/r_s}$, and timescale $t_0^{-1}=a \sigma_m v_0 \rho_s$.
There are thus two main inputs: the initial density profile $\rho_{\mathrm{init}}$ and the SIDM cross section per mass $\sigma_m$.

The first step is the discretization of the spherical halo.
We divide the halo into $N$ concentric shells, and the outer radii of the shells are logarithmically spaced between $\tilde{r}_{\min} = \tilde{r}_1$ and $\tilde{r}_{\max} = \tilde{r}_{N}$.
In this work, we choose $\tilde{r}_{\min} = 0.01$, $\tilde{r}_{\max} = 100$, and $N = 400$.

Extensive variables associated with the $i$-th shell are defined at $\tilde{r}_i$: $\tilde{M}_i$ and $\tilde{V}_i$ are the dimensionless mass and volume contained within the radius $\tilde{r}_i$, and $\tilde{L}_i$ is the dimensionless luminosity at $\tilde{r}_i$.

Intensive variables of the $i$-th shell are the dimensionless density $\tilde{\rho}_i$, pressure $\tilde{p}_i$, specific energy $\tilde{u}_i$, and velocity dispersion $\tilde{v}_i$; for concreteness, these quantities are defined at the midpoint of the $i$-th shell, $(\tilde{r}_i + \tilde{r}_{i-1})/2$.
They are also related to one another through the equipartition theorem and ideal gas law: $\tilde{p}_i = \tilde{\rho}_i \tilde{v}_i^2 = \frac{2}{3} \tilde{\rho}_i \tilde{u}_i$.

From the input initial density, it is straightforward to obtain $\tilde{M}_i$ and $\tilde{p}_i$ from Eqs.~\eqref{eq:mass_conservation} and \eqref{eq:hydrostatic}, respectively.
Given a value of $\tilde{\sigma}_m$, we can determine $\tilde{L}_i$ from Eq.~\eqref{eq:thermo}.
With all quantities initialized, we evolve the halo by taking a small heat conduction time evolution step, followed by adjustments to maintain hydrostatic equilibrium.

\begin{figure*}[t]
  \includegraphics[width=0.329\textwidth]{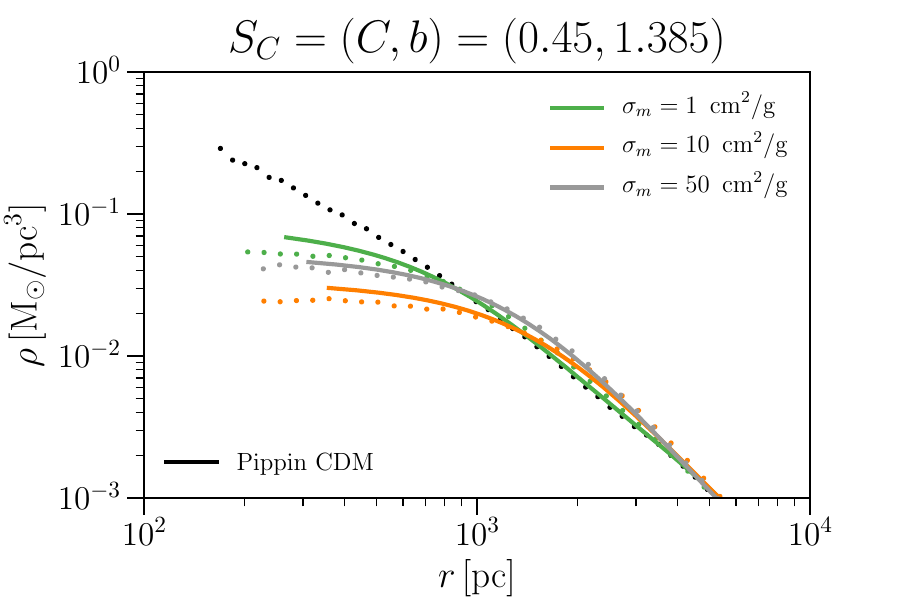}
  \includegraphics[width=0.329\textwidth]{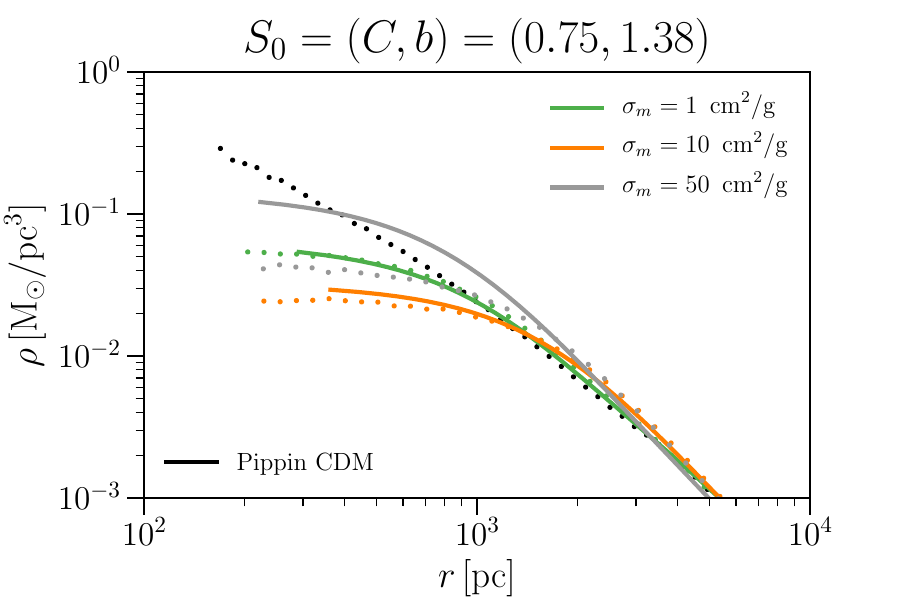}
  \includegraphics[width=0.329\textwidth]{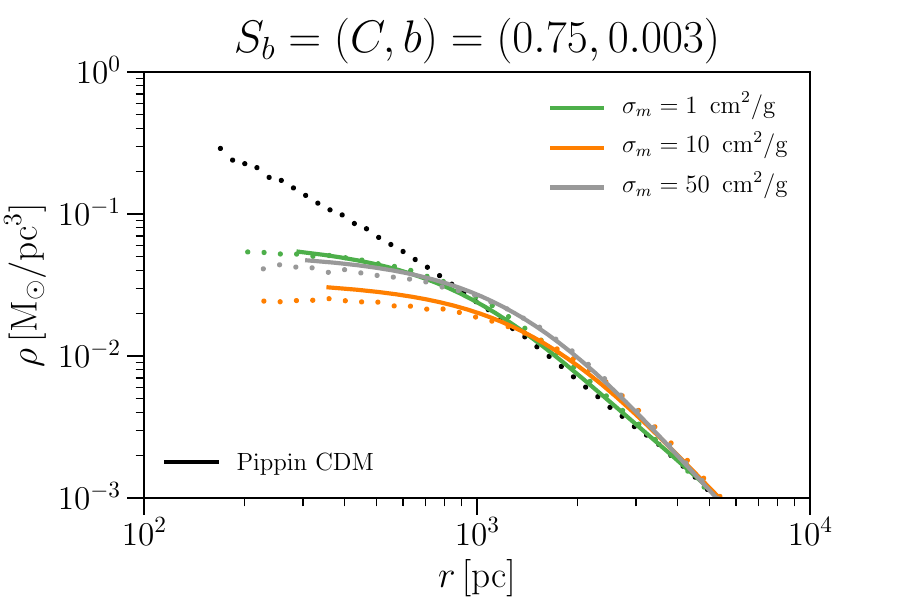}\\
  \includegraphics[width=0.329\textwidth]{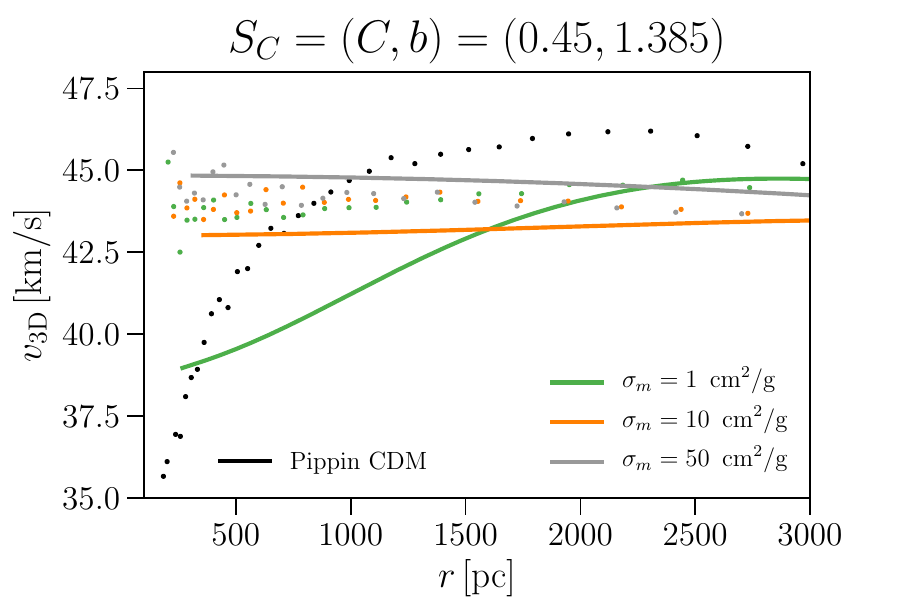}
  \includegraphics[width=0.329\textwidth]{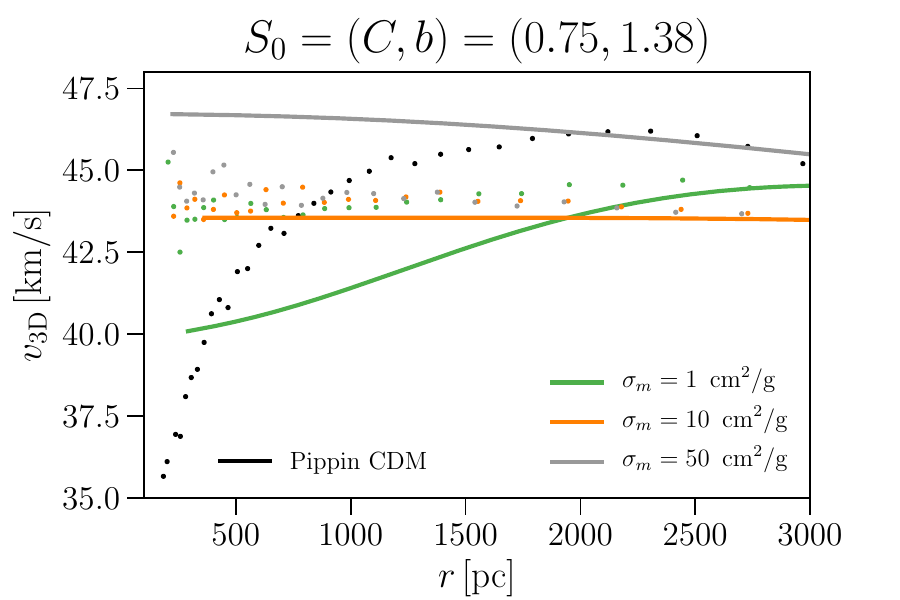}
  \includegraphics[width=0.329\textwidth]{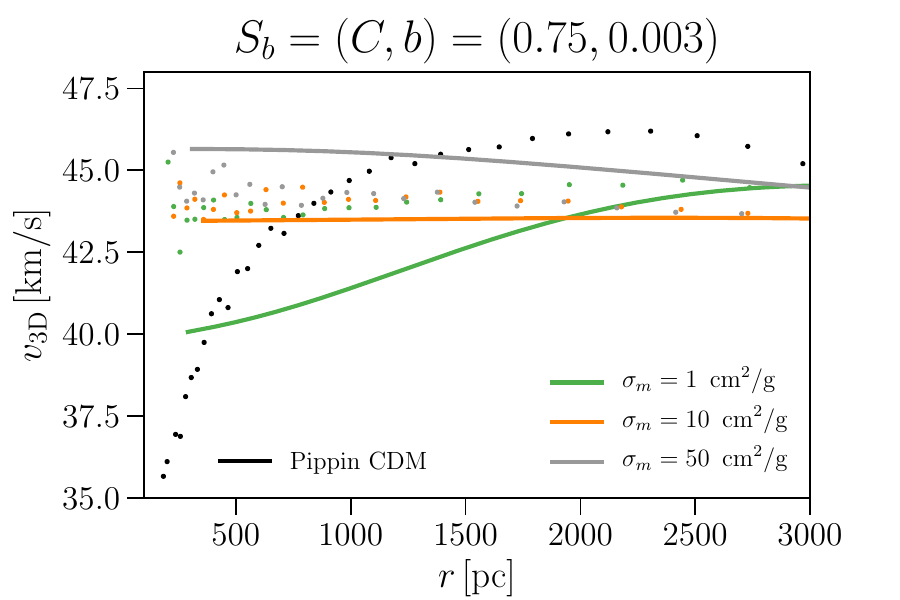}
  \caption{Calibrations of $C$ and $b$ using the density profiles (top panels) and 3D velocity dispersions (bottom panels) of the Pippin halo in Ref.~\cite{Elbert:2014bma}.
    Reference~\cite{Essig:2018pzq} chose the parameters $S_C=(C, b)=(0.45, 1.38)$ (left panels) to match (using an initially NFW profile) the $\sigma_m=50~\cm^2/\g$ density curve of Pippin by adjusting the LMFP heat transfer parameter $C$.
    Our results use $S_0=(0.75, 1.38)$ (middle panels), which provides a good fit to the density profiles for $\sigma_m \leq 10~\cm^2/\g$.
    Calculations with $S_b=(0.75, 0.003)$ (right panels) lower the density profile for $\sigma_m=50~\cm^2/\g$, while maintaining agreement with Pippin for smaller $\sigma_m$ values.}
  \label{fig:Pippin_Calibration}
\end{figure*}

To perform the heat conduction time step, we rewrite Eq.~\eqref{eq:thermo} as
\begin{equation}
  \delta\tilde{u} \simeq -\left(\frac{\partial \tilde{L}}{\partial \tilde{M}}\right) \delta \tilde{t},
  \label{eq:du}
\end{equation}
where the change in density is taken to be negligible (such that the mass within the shells remains constant) for a sufficiently small time step $\delta \tilde{t}$.
This condition is ensured throughout the calculation by imposing $\delta \tilde{t}$ to be smaller than the minimum local relaxation time of the halo by $\epsilon_t (\ll 1)$.
In particular, we require $\delta\tilde{t} = \epsilon_t \min_{0\leq i \leq N} (t_{r,i}/t_0)$, where $t_{r, i}$ is the local relaxation time of the $i$-th shell, and we set $\epsilon_t = 10^{-4}$.

The heat conduction causes an increase in the specific energy, which pushes the halo out of hydrostatic equilibrium.
With the assumption of negligible change in $\tilde{\rho}_i$ in this step, the increase in $\tilde{u}_i$ is reflected in the following step through $\tilde{p}_i = \frac{2}{3}\tilde{\rho}_i\tilde{u}_i$.
To return the halo to hydrostatic equilibrium, we adjust the shell location $\tilde{r}_i \to \tilde{r}'_i \equiv \tilde{r}_i + d \tilde{r}_i$, pressure $\tilde{p}_i \to \tilde{p}'_i \equiv \tilde{p}_i + d \tilde{p}_i$, and density $\tilde{\rho}_i \to \tilde{\rho}'_i \equiv \tilde{\rho}_i + d \tilde{\rho}_i$.

The conservation of mass $M_i$ enforces $d \tilde{\rho}_i = -\tilde{\rho}_i (d\tilde{V}_i/\tilde{V}_i)$, where $d\tilde{V}_i$ is the change in volume associated with the shift in radii.
The conservation of entropy requires $\tilde{p}_i \tilde{V}^{5/3}_i$ to be constant after the adjustment.
Through this process, the discretized form of the hydrostatic condition becomes
\begin{equation}
  \frac{\tilde{p}'_{i+1}-\tilde{p}'_i}{(\tilde{r}'_{i+1}-\tilde{r}'_{i-1})/2}
  = - \frac{\tilde{M}_{i}}{\tilde{r}_i^{\prime 2}}
  \frac{\tilde{\rho}'_{i+1}+\tilde{\rho}'_{i}}{2}
    \label{eq:adjustment}
\end{equation}
for $i = 1,\dots, N-1$.
We define a fixed inner boundary shell $\tilde{r}_0 = 0$ with $\tilde{L}_0=0$ (i.e., no source of heat at the center of the halo), and we fix the boundary of the outermost shell such that $d\tilde{r}_{N}=0$.

When linearized, Eq.~\eqref{eq:adjustment} takes up the form of $a_i\, d\tilde{r}_{i-1}+b_i\, d\tilde{r}_{i}+c_i\, d\tilde{r}_{i+1}=d_i$, where $a_i$, $b_i$, $c_i$, and $d_i$ are constants.
The result is a set of $N-1$ tridiagonal equations for $N-1$ variables $d\tilde{r}_1, \dots, d\tilde{r}_{N-1}$.
After solving this system of equations, we update all other variables.
The hydrostatic adjustment is repeated in between every heat conduction step until $\max_{0< i < N} |d\tilde{r}_i/\tilde{r}_i| < \epsilon_r$ is satisfied, where we set $\epsilon_r = 10^{-3}$.

\section{Calibration}
\label{app:calibration}
In the gravothermal fluid model, the two parameters $C$ and $b$ need to be calibrated against N-body simulations, as they adjust the efficiency of the LMFP and SMFP heat transfer process, respectively, in Eq.~\eqref{eq:gravothermal}.
Reference~\cite{Koda:2011yb} matched the collapse time of pure LMFP evolution (i.e., ${\sigma_m}^2 / b \to 0$) to first calibrate $C$ for the case of an initially self-similar solution of an isolated halo.
The value of $b$ was then decreased from $b=1.38$ to $0.25$ to match the collapse times of self-similar halos with larger values of $\sigma_m$.
Instead of matching the collapse times, we determine the values of $C$ and $b$ by directly comparing the calculated density profiles after $13~\mathrm{Gyr}$ of evolution against the corresponding profiles of the Pippin halo, generated from a cosmological simulation~\cite{Elbert:2014bma}.

Figure~\ref{fig:Pippin_Calibration} shows the results of SIDM evolutions with different $(C, b)$ values: $S_0=(0.75, 1.38)$, $S_C = (0.45, 1.38)$, and $S_b = (0.75, 0.003)$.
Used in Ref.~\cite{Essig:2018pzq}, $S_C$ provides a better match of the density profile for $\sigma_m=50~\cm^2/\g$ but worsens agreement for lower $\sigma_m$.
However, the Pippin halo with $\sigma_m=50~\cm^2/\g$ exhibits signs of core collapse, so we do not use it to calibrate $C$, which controls LMFP heat transfer.
Instead, we calibrate $C$ against the Pippin halo for cross sections $\sigma_m < 50~\cm^2/\g$, keeping $b=1.38$ fixed.
The profiles with $S_0$ in Fig.~\ref{fig:Pippin_Calibration} show good agreement for $\sigma_m \leq 10~\cm^2/\g$, but a clear discrepancy for $\sigma_m = 50~\cm^2/\g$.
We note that the results for $S_0$ and $S_C$ are essentially the same for the key points in this paper.

In the spirit of Ref.~\cite{Koda:2011yb}, we allow for adjustments in $b$ and find that $b=0.003$ lowers the density profile with $\sigma_m = 50~\cm^2/\g$.
This solution, however, never reduces the density as low as the $10~\cm^2/\g$ case requires, and it pushes the early evolution into the SMFP regime.
In addition, the dispersion profile for $10~\cm^2/\g$ (as well as $\sigma_m = 5~\cm^2/\g$) does not have the negative radial gradient at $t=13~\Gyr$ that the simulation exhibits.

In light of the discussion above, we see no compelling reason to adopt the $S_b$ solution.
Instead, the key issue is likely that the heat conduction $\kappa^{-1} = {\kappa_\textrm{SMFP}}^{-1} + {\kappa_\textrm{LMFP}}^{-1}$ does not correctly capture the physics of the transition between the SMFP and LMFP regimes.
We could generalize this interpolation to $\kappa^{-\alpha} = {\kappa_\textrm{LMFP}}^{-\alpha} +{\kappa_\textrm{SMFP}}^{-\alpha}$, such that
\begin{equation}
  \tilde{L} = -\frac{3C}{2} {\tilde{r}}^2 \tilde{\rho} {\tilde{v}}^3
        {(1+x^{\alpha})}^{-1/\alpha}
        \frac{\partial {\tilde{v}}^2}{\partial \tilde{r}} \ ,
\end{equation}
where $x \equiv (aC/b){\tilde{\sigma}_m}^2 \tilde{\rho} {\tilde{v}}^2$.
This generalization would allow us to increase $C$ and tune $\alpha$ to obtain a faster evolution, while decreasing the impact of SMFP regime.
We leave such an investigation for future work.

\begin{figure}[t]
  \includegraphics[width=\columnwidth]{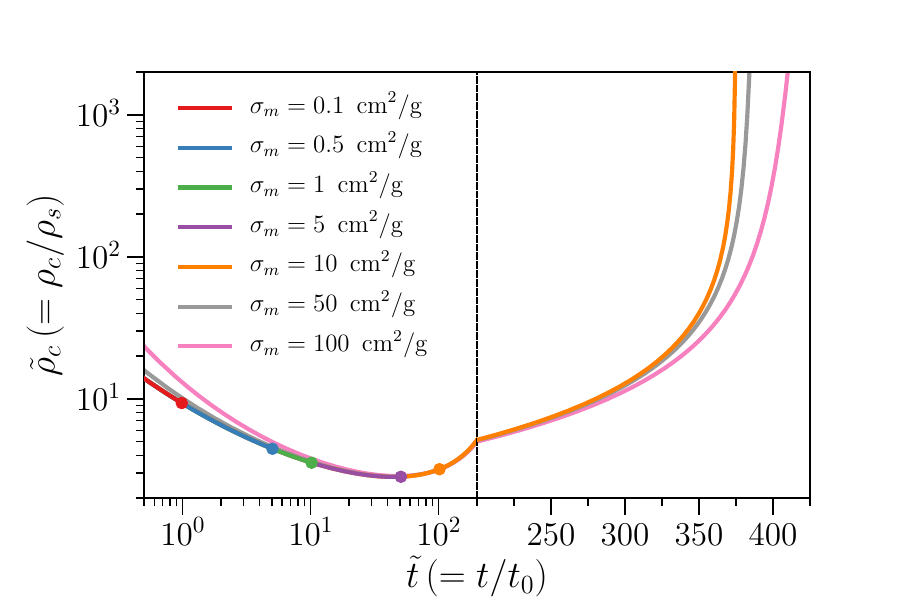}
  \caption{Central density evolution of an initially NFW profile.
    The evolution for $\sigma_m \leq 10~\cm^2/\g$ is driven by LMFP heat conduction for $\tilde{t} \leq 374$, and the curves align on top of one another.
    SMFP heat transfer with $\sigma_m \geq 50~\cm^2/\g$ is non-negligible near the beginning and the end of the evolution, as indicated by the deviations from the curves for $\sigma_m \leq 10~\cm^2/\g$.
    The dots correspond to $t=13~\Gyr$ for each $\sigma_m$ (dots for $\sigma_m \geq 50~\cm^2/\g$ are outside the plotted range).}
  \label{fig:central_density_deviation}
\end{figure}

\section{Halo evolution}
\label{app:evolution}
In Fig.~\ref{fig:central_density_deviation}, we show the time evolution of the halo central density with various values of $\sigma_m$ for an initially NFW profile.
We use the dimensionless time $\tilde{t}=t/t_0$, so larger values of $\sigma_m$ correspond to longer dimensionless times.
For the cross section $\sigma_m = 10~\cm^2/\g$, we evolve the central density beyond $13~\Gyr$ to show the gravothermal catastrophe as the central density sharply rises at $\tilde{t} \gtrsim 350$.

There are a few important features to note.
First, the curves for $\sigma \leq 5~\cm^2/\g$ align with the $\sigma_m = 10~\cm^2/\g$ curve when extended, indicating that the evolution for $\sigma_m \leq 10~\cm^2/\g$ occurs in the LMFP regime at times of interest and is insensitive to the value of $\sigma_m$.
Second, the deviations of the $50$ and $100~\cm^2/\g$ curves away from the lower $\sigma_m$ curves at small values of $\tilde{t}$ indicate that SMFP physics is not negligible at these early times.
At late times, the large $\sigma_m$ curves both bifurcate from the self-similar solution.
Finally, for all values of $\sigma_m$ we tested, there is a minimal central density of $\tilde{\rho}_c = 2.8$ that the SIDM halo achieves at $\tilde{t} \simeq 45$.

\begin{figure*}[t]
  \includegraphics[width=0.329\textwidth]{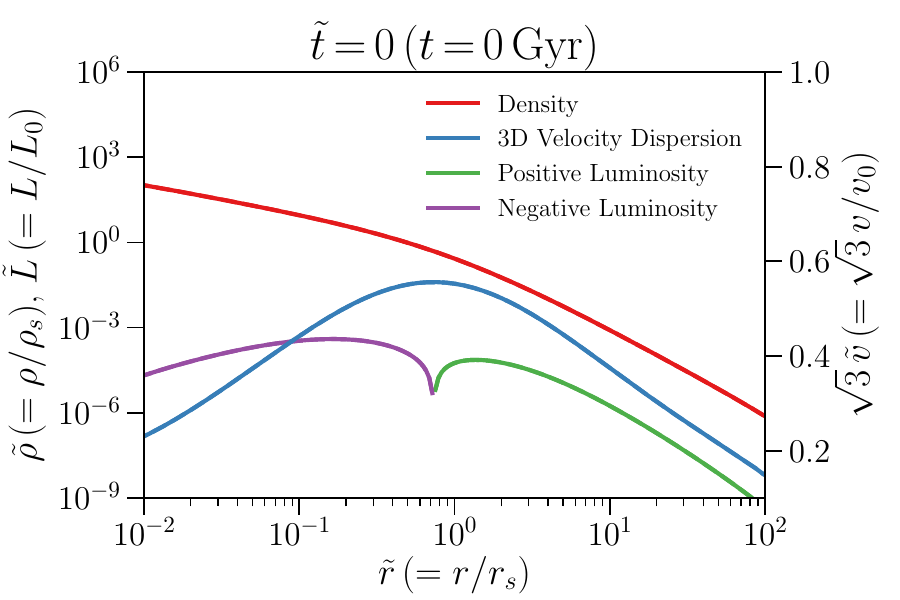}
  \includegraphics[width=0.329\textwidth]{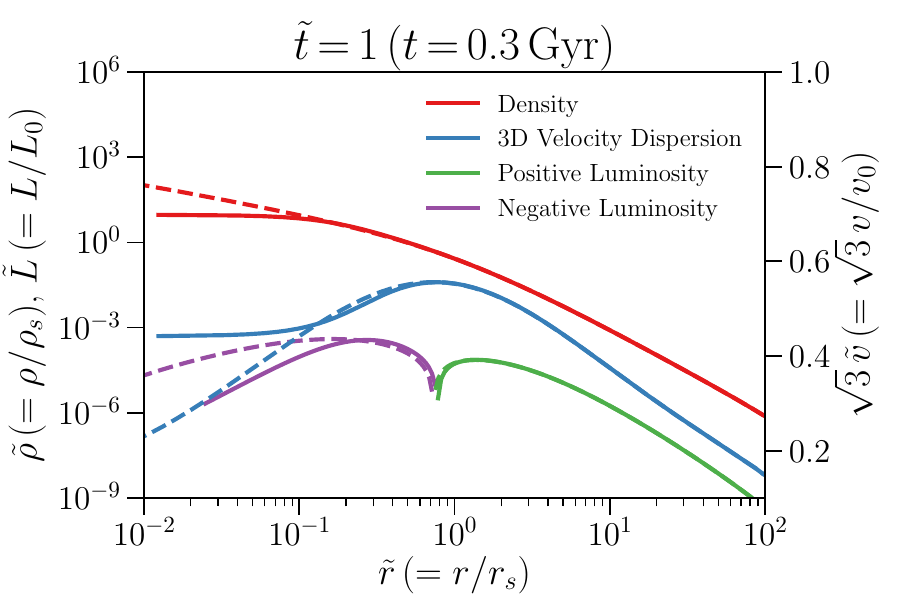}
  \includegraphics[width=0.329\textwidth]{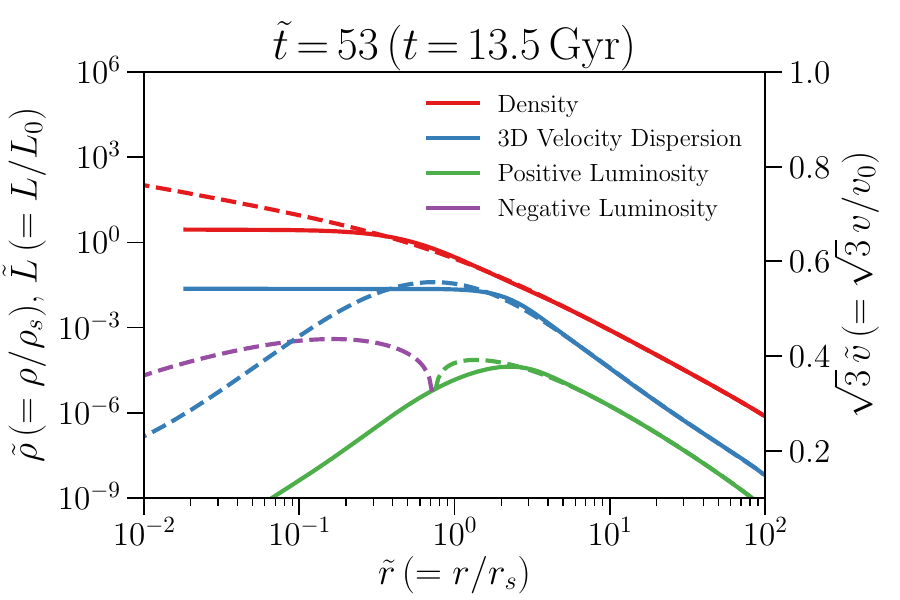}\\
  \includegraphics[width=0.329\textwidth]{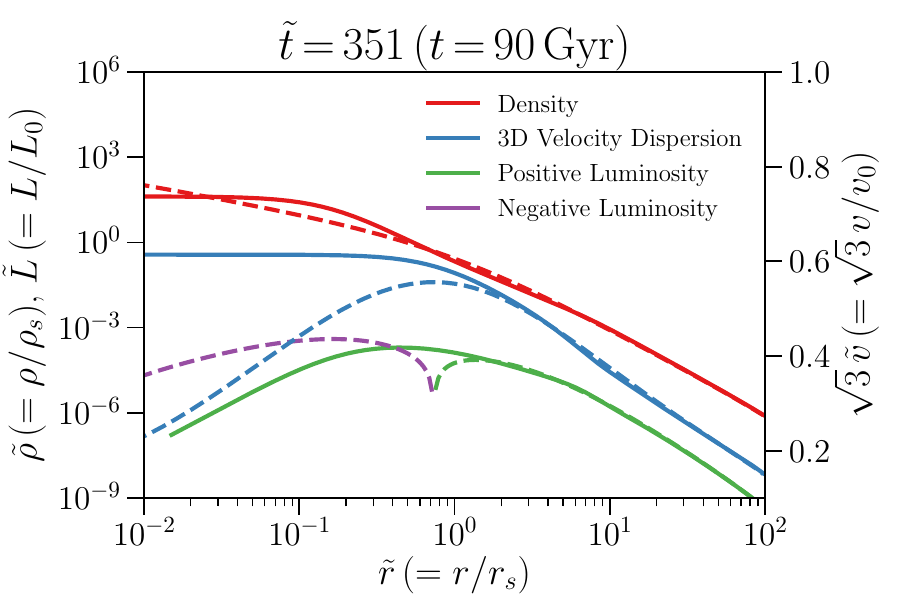}
  \includegraphics[width=0.329\textwidth]{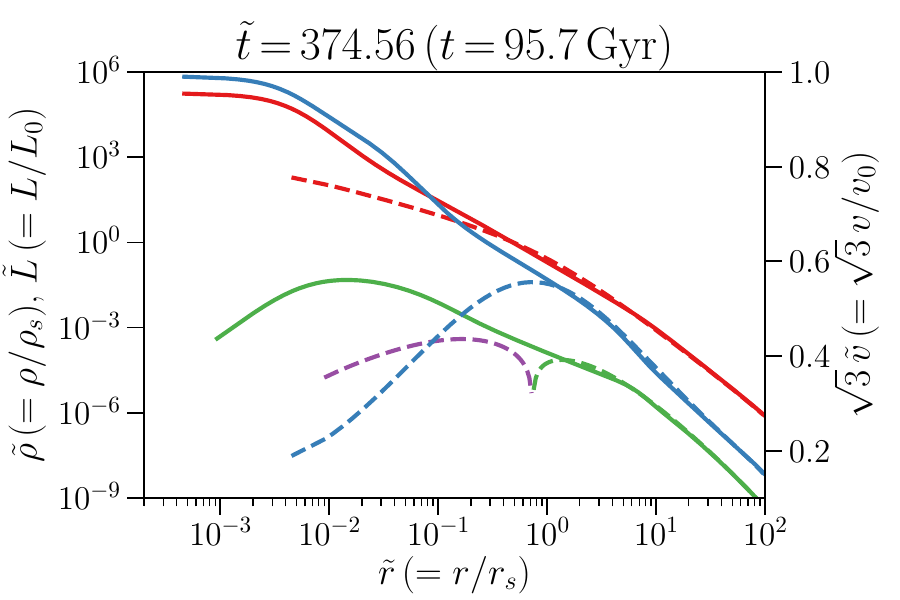}
  \caption{Snapshots of the halo evolution for an initially NFW profile with cross section per mass $\sigma_m=5~\cm^2/\g$ at times $\tilde{t} = 0$, $1$, $53$, $351$, and $374.56$.
    The density profile and luminosity are shown on a log-log scale as functions of the radius.
    The 3D velocity dispersion is plotted on a linear-log scale, with the linear axis given on the right.
    All the quantities are represented as dimensionless variables, defined below Eq.~\eqref{eq:gravothermal}, with $v_0 \simeq 84~\km/\s$, $t_0 \simeq 0.255~\Gyr$, and $L_0 \simeq 1.9 \times 10^7\, L_\odot$ for $\rho_s = 0.019~\Msun/\pc^3$ and $r_s = 2.59~\kpc$.}
  \label{fig:snapshots_NFW}
\end{figure*}

To demonstrate the formation of the core and its subsequent collapse, we consider the evolution of an initially NFW halo with $\sigma_m = 5~\cm^2/\g$.
Figure~\ref{fig:snapshots_NFW} shows snapshots of the halo density profile, 3D velocity dispersion, and luminosity at 5 different times.
We choose the same calibration $S_0 = (C, b) = (0.75, 1.38)$ and halo parameters ($\rho_s = 0.019~\Msun/\pc^3$ and $r_s = 2.59~\kpc$) used in the main text.
The state of the halo in each panel of Fig.~\ref{fig:snapshots_NFW} is as follows:
\begin{enumerate}
\item $\tilde{t} = 0$: Initial NFW profile.
  The initial profile satisfies $\kappa_{\mathrm{SMFP}} \geq \kappa_{\mathrm{LMFP}}$ throughout the halo for $\sigma_m \lesssim 10~\cm^2/\g$ (cf.\ Fig.~\ref{fig:central_density_deviation}).
  In fact, heat transfer remains dominated by the LMFP physics until the core of the halo becomes sufficiently dense at $\tilde{t} \simeq 374$.
\item $\tilde{t} = 1$: Core expansion.
  Self-interactions thermalize the inner region of the halo, where the mean-free path $\lambda = 1/(\rho\sigma_m)$ is short.
  As a result, the temperature ($\propto {\tilde{v}}^2$) near the center begins to increase, and the particles are pushed towards $r\sim r_s$.
  This leads to the formation of a gradually expanding core.
\item $\tilde{t} = 53$: Formation of the isothermal core.
  The luminosity becomes entirely positive (shortly after a minimum central density of $\tilde{\rho}_c = 2.8 \, \rho_s$ is achieved at $\tilde{t} = 45$) after a period of core expansion, resulting from the thermalization within the innermost region of the halo.
  The core becomes isothermal, as $\tilde{v}$ is now nearly constant within this region.
  This time marks the turnaround of the central density evolution, seen in the left panel of Figs.~\ref{fig:central_density_evolution} and \ref{fig:central_density_deviation}.
\item $\tilde{t} = 351$: LMFP core collapse.
  The constant-density core becomes denser as it shrinks in size, while the density profile maintains its shape since the evolution is self-similar.
  The temperature uniformly increases within the core such that it remains isothermal.
  This phase corresponds to the solid lines in Fig.~\ref{fig:rho_core_temperature}.
\item $\tilde{t} = 374.56$: SMFP core collapse.
  The condition $\kappa_{\mathrm{SMFP}} \geq \kappa_{\mathrm{LMFP}}$ is satisfied entirely within the core, and the halo central density deviates from the self-similar solution.
  At the time shown, the SMFP core has attained its maximum mass.
  The density and temperature begins to increase more rapidly within the SMFP core as the core becomes optically thick to self-interactions.
  This final phase of evolution is shown as dashed lines in Fig.~\ref{fig:rho_core_temperature}; the core retains a substantial amount of its mass until it reaches the relativistic instability to form a BH\@.
\end{enumerate}

\begin{figure*}[t]
  \includegraphics[width=0.329\textwidth]{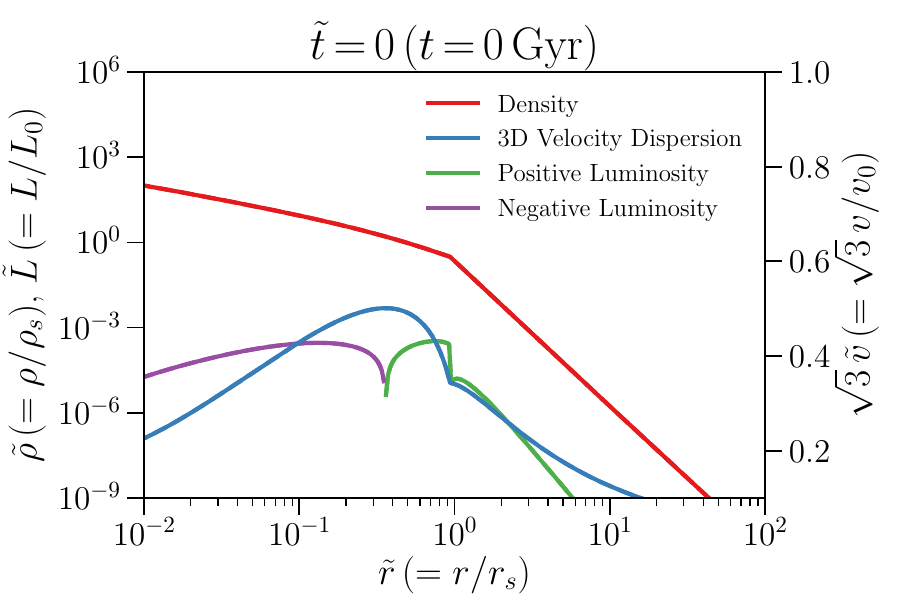}
  \includegraphics[width=0.329\textwidth]{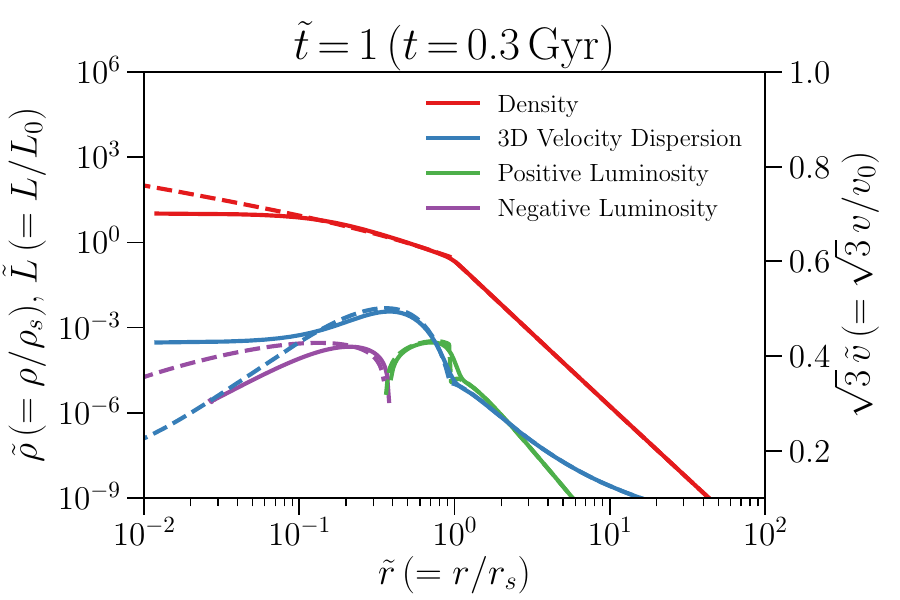}
  \includegraphics[width=0.329\textwidth]{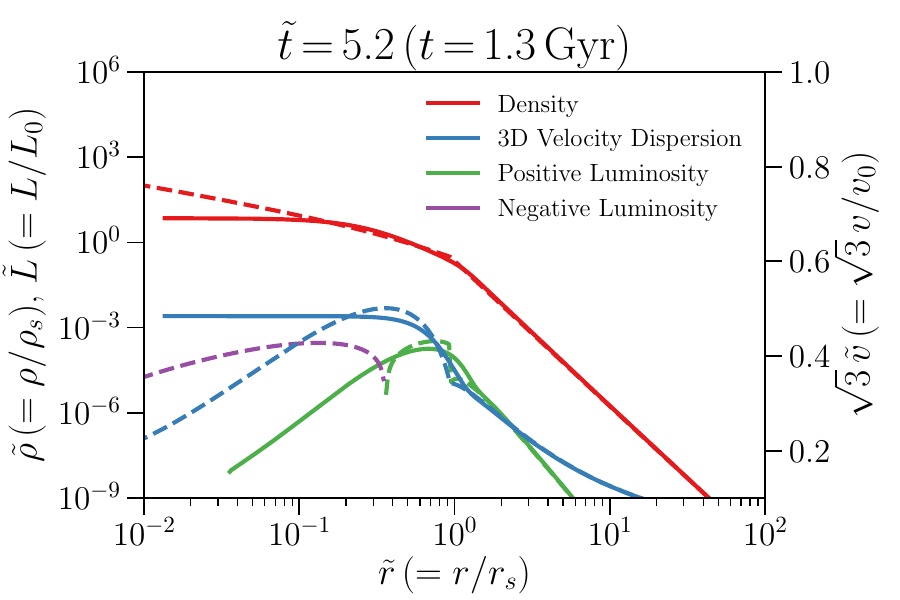}\\
  \includegraphics[width=0.329\textwidth]{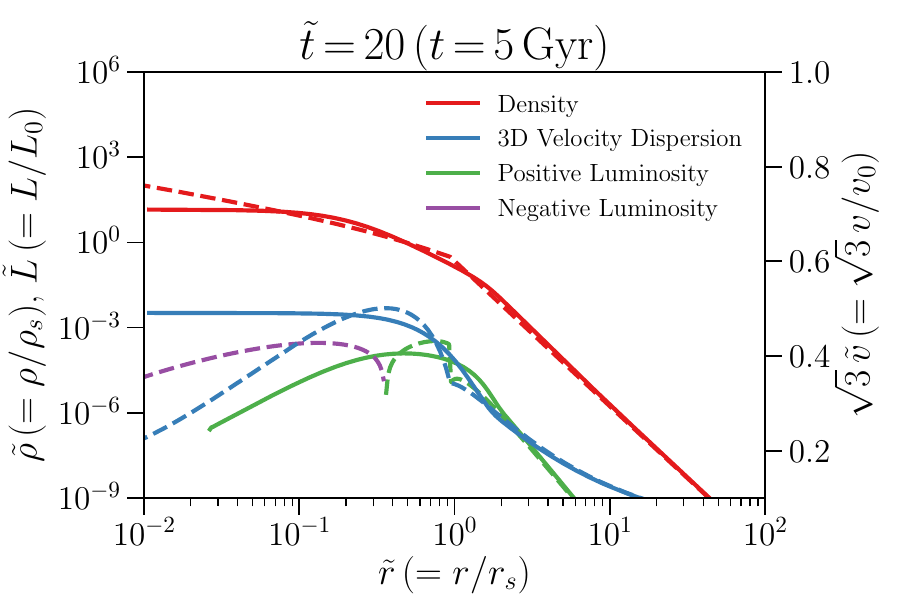}
  \includegraphics[width=0.329\textwidth]{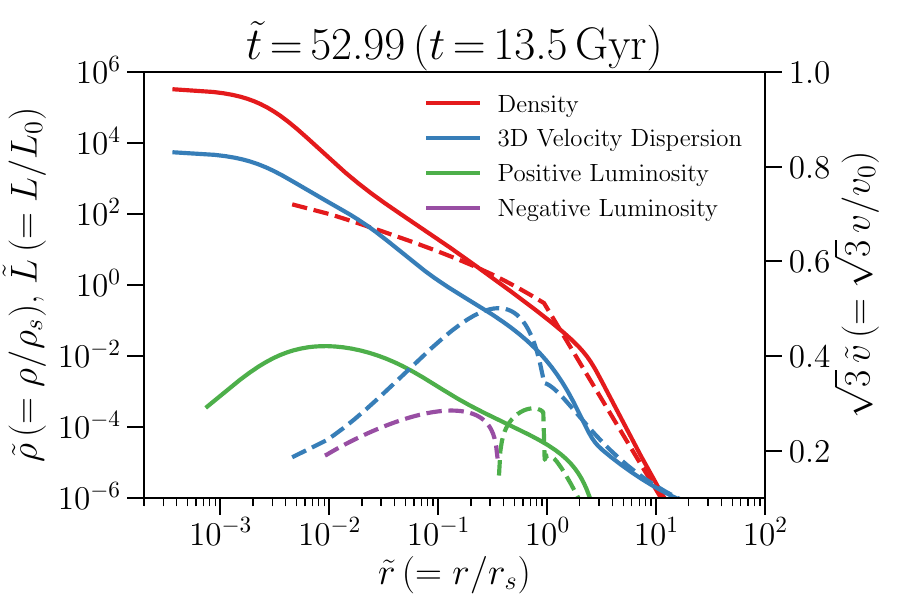}
  \caption{Same as Fig.~\ref{fig:snapshots_NFW}, but for a halo with an initially truncated NFW profile with $r_t=r_s$ at $\tilde{t} = 0$, $1$, $5.2$, $20$, and $52.99$.}
  \label{fig:snapshots_TNFW}
\end{figure*}

\begin{figure*}[t]
  \includegraphics[width=0.329\textwidth]{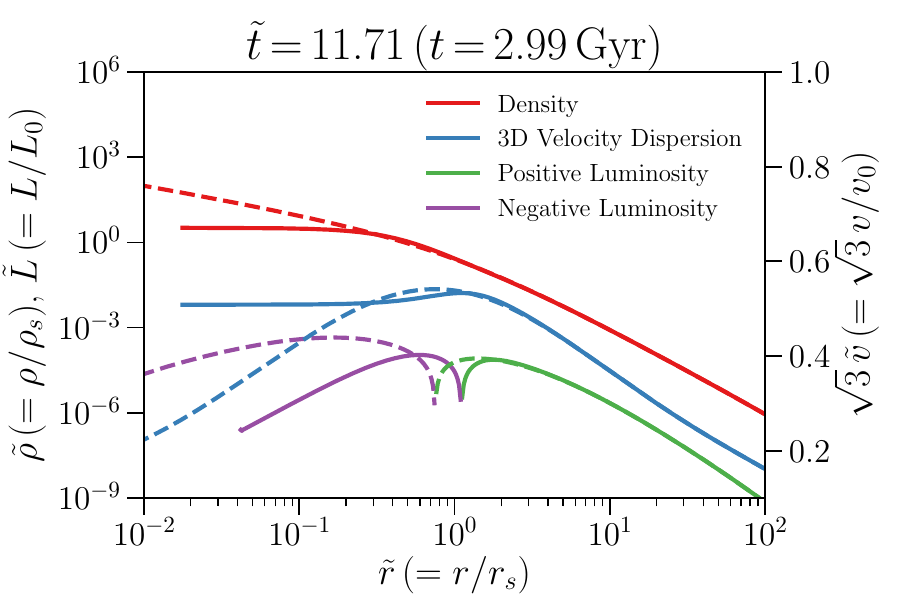}
  \includegraphics[width=0.329\textwidth]{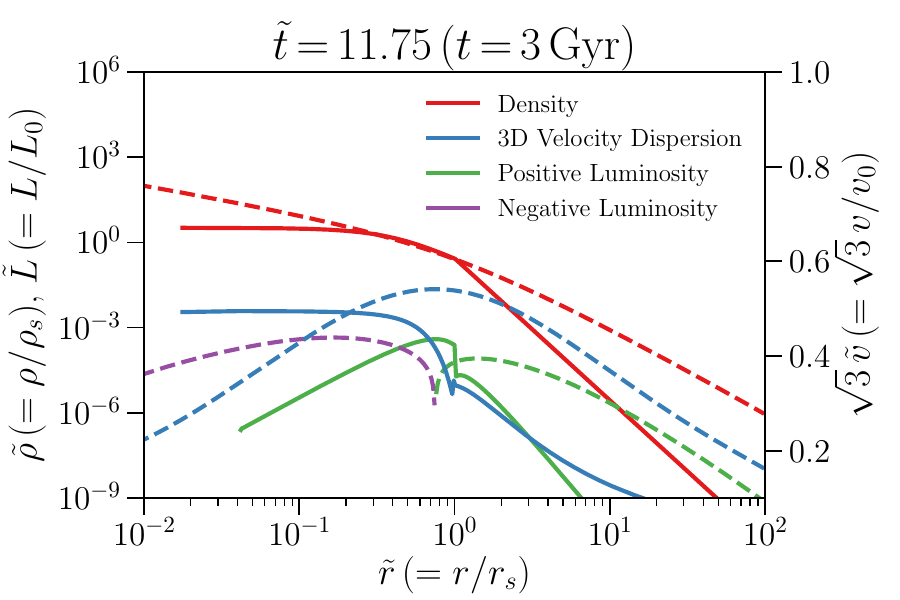}
  \includegraphics[width=0.329\textwidth]{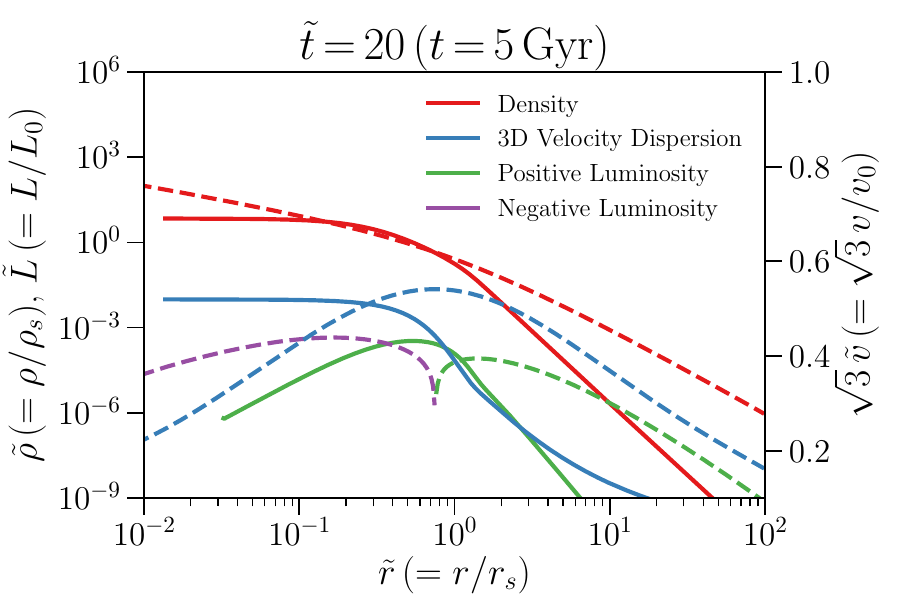}\\
  \includegraphics[width=0.329\textwidth]{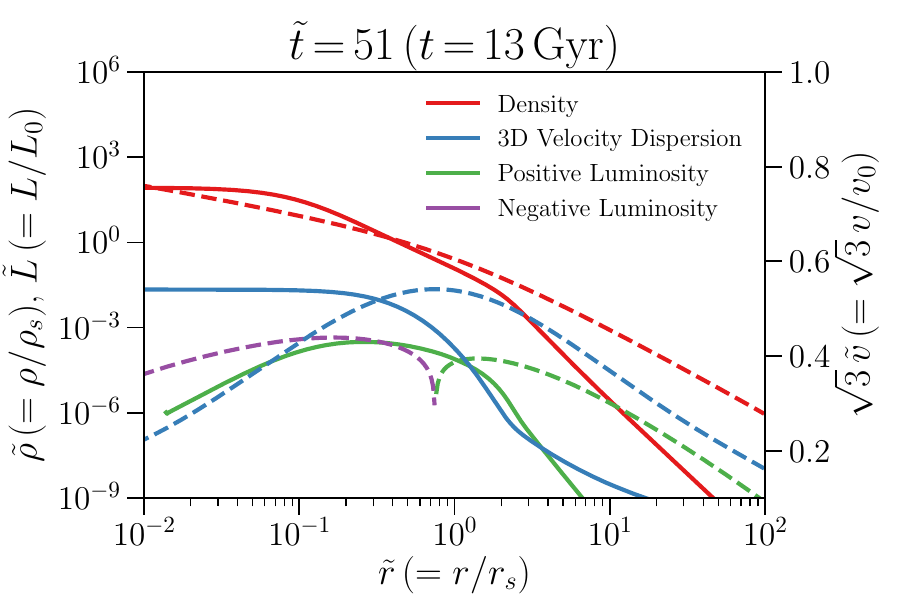}
  \includegraphics[width=0.329\textwidth]{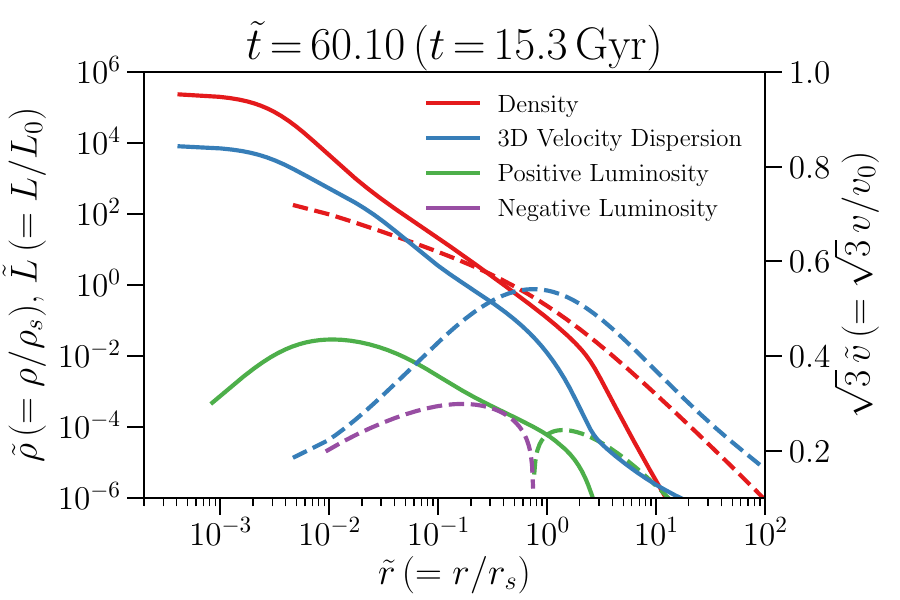}
  \caption{Same as Fig.~\ref{fig:snapshots_NFW}, but for a halo with an initially NFW profile, truncated at $r_t=r_s$ after $3~\Gyr$ of evolution, at $\tilde{t} = 11.71$, $11.75$, $20$, $51$, and $60.10$.
    The snapshots at $\tilde{t} = 11.71$ and $11.75$ correspond to before and after truncation.}
   \label{fig:snapshots_TNFW3Gyr}
\end{figure*}

Figure~\ref{fig:snapshots_TNFW} shows analogous snapshots for the evolution of an initially TNFW profile at $r_t=r_s$ with $\sigma_m = 5~\cm^2/\g$.
We describe the state of the halo in each panel in the following:
\begin{enumerate}
\item $\tilde{t}=0$: Initial TNFW profile.
  The outer region of the halo ($r \geq r_s$) is assumed to be tidally stripped upon formation of an NFW halo, and the density is suppressed by a factor of $(r/r_t)^5$ beyond the truncation radius.
  The truncation reduces the pressure of the halo beyond the truncation radius, due to enforcing hydrostatic equilibrium.
  Although the density is decreased beyond the truncation radius as well, the drop in pressure is more drastic, causing the velocity dispersion (or equivalently, temperature) profile to decrease as well.
  Thus, the outer part of the halo gets significantly colder, and the temperature peak around $r \sim r_s$ becomes smaller and narrower compared to the NFW profile without truncation.
\item $\tilde{t}=1$: Core expansion.
  Compared to the initially NFW halo in Fig.~\ref{fig:snapshots_NFW} at the same time, the core is closer to being fully thermalized due to the less pronounced peak in velocity dispersion.
\item $\tilde{t}=5.2$: Formation of the isothermal core.
  The luminosity becomes entirely positive, and the core expansion halts at a much earlier time than seen in Fig.~\ref{fig:snapshots_NFW}.
  As a result, the isothermal core is more concentrated: its size is smaller, and its density is higher ($\rho_c \simeq 7.1 \, \rho_s$).
\item $\tilde{t}=20$: LMFP core collapse.
  The core contracts slowly in the same way as seen for the initially NFW halo in Fig.~\ref{fig:snapshots_NFW} through the LMFP heat transport.
\item $\tilde{t}=52.99$: SMFP core collapse.
  The halo reaches the same state as the initially NFW halo at $\tilde{t}=374.56$.
  Collapse occurs $13.5~\Gyr$ after halo formation, implying that such a truncation causes a gravothermal catastrophe with $\sigma_m = 5~\cm^2/\g$ on timescales relevant for what we may observe today.
\end{enumerate}

Lastly, we show in Fig.~\ref{fig:snapshots_TNFW3Gyr} analogous snapshots for the evolution of an initially NFW halo that is truncated at $r_t=r_s$ after $3~\Gyr$ with $\sigma_m = 5~\cm^2/\g$.
We describe the state of the halo in each panel in the following:
\begin{enumerate}
\item $\tilde{t} = 11.71$: Before truncation.
  The halo follows the evolution of the initially NFW profile (the first two panels in Fig.~\ref{fig:snapshots_NFW} are the same for this case).
  The negative luminosity within $r \leq r_s$ indicates that the core is gradually expanding.
\item $\tilde{t} = 11.75$: After truncation.
  Compared to the initially TNFW halo in Fig.~\ref{fig:snapshots_TNFW}, the (instantaneous) late truncation at $\tilde{t}=11.75$ ($t=3~\Gyr$) results in a slightly different halo configuration.
  When the late truncation lowers the temperature of the halo beyond the truncation radius, the already well-thermalized core immediately becomes hotter than the $r \sim r_s$ region where the temperature used to peak.
  This flips the sign of the luminosity within the core from negative to positive and triggers the core contraction phase seen in the initially NFW halo after $\tilde{t}=53$ in Fig.~\ref{fig:snapshots_NFW}.
\item $\tilde{t} = 20$: $2~\Gyr$ after truncation.
  The halo properties have smoothed out after the sudden truncation, and the core is gradually contracting under the LMFP evolution.
  Although the central density is lower than the initially TNFW profile at $\tilde{t}=20$, the central velocity dispersion is higher due to the initial thermalization phase prior to truncation.
\item $\tilde{t} = 51$: $13~\Gyr$ after the initial formation of the NFW halo.
  While the initially NFW halo reaches the minimum core density around the same time (panel 3 of Fig.~\ref{fig:snapshots_NFW}), the core of the truncated halo has a density that is $\sim\! 30$ times higher.
\item $\tilde{t} = 60.10$: SMFP core collapse.
  At $15.3~\Gyr$, the innermost region of the halo is undergoing the same collapse process that its NFW counterpart reaches at $95.7~\Gyr$ after its formation.
\end{enumerate}

Finally, in Table~\ref{table:SMFP_core}, we show the values of the mass of the SMFP core $M_{\textrm{SMFP}}$ and the 3D velocity dispersion at the center of the halo $v_{3 \mathrm{D}, c}$ when $M_{\textrm{SMFP}}$ reaches the maximum value.
In the main text, we used these values in the empirical relation~\cite{Balberg:2002ue} to estimate the mass of BH that forms in the gravothermal collapse in the SMFP regime for the case of $\sigma_m = 5~\cm^2/\g$.
We note that the last panels of Figs.~\ref{fig:snapshots_NFW},~\ref{fig:snapshots_TNFW}, and~\ref{fig:snapshots_TNFW3Gyr} show the halo profiles at these times, the central regions of which have all deviated from the self-similar evolution.

\begingroup
\setlength{\tabcolsep}{12pt}
\renewcommand{\arraystretch}{1.5}
\begin{table*}[t]
  \centering
  \begin{tabular}{c c c c c c c}
    Profile & $\tilde{t}_f$ & $t_f \ [\Gyr]$ & $M_{\mathrm{SMFP}}(\tilde{t}_f) \ [10^7 \, \Msun]$ & $v_{3\textrm{D}, c}(\tilde{t}_f) \ [\km/\s]$\\
    \hline \hline
    NFW & 374.56 & 95.66 & 5.7 & 83\\
    \hline
    TNFW $r_t=3\,r_s$ & 183.39 & 46.84 & 4.6 & 81\\
    \hline
    TNFW $r_t=r_s$ after $3 \, \Gyr$ & 60.10 & 15.34 & 3.0 & 59\\
    \hline
    TNFW $r_t=r_s$ & 52.99 & 13.53 & 2.6 & 61\\
  \end{tabular}
  \caption{The core mass and 3D velocity dispersion for various profiles with $\sigma_m = 5~\cm^2/\g$ at the time of maximum SMFP-core formation $\tilde{t}_f$, roughly when the condition $\kappa_{\mathrm{SMFP}} \leq \kappa_{\mathrm{LMFP}}$ becomes true over the entire core region.
    Note that this definition of the core mass $M_{\mathrm{SMFP}}$ differs from the core mass $M_\textrm{core}$ used in Fig.~\ref{fig:rho_core_temperature}.
    We use $M_{\mathrm{SMFP}}$ when extrapolating the empirical relation found in Ref.~\cite{Balberg:2002ue}}
  \label{table:SMFP_core}
\end{table*}
\endgroup

\bibliographystyle{apsrev4-1.bst}
\bibliography{sidmhaloBib}

\begin{thebibliography}{51}%
\makeatletter
\providecommand \@ifxundefined [1]{%
 \@ifx{#1\undefined}
}%
\providecommand \@ifnum [1]{%
 \ifnum #1\expandafter \@firstoftwo
 \else \expandafter \@secondoftwo
 \fi
}%
\providecommand \@ifx [1]{%
 \ifx #1\expandafter \@firstoftwo
 \else \expandafter \@secondoftwo
 \fi
}%
\providecommand \natexlab [1]{#1}%
\providecommand \enquote  [1]{``#1''}%
\providecommand \bibnamefont  [1]{#1}%
\providecommand \bibfnamefont [1]{#1}%
\providecommand \citenamefont [1]{#1}%
\providecommand \href@noop [0]{\@secondoftwo}%
\providecommand \href [0]{\begingroup \@sanitize@url \@href}%
\providecommand \@href[1]{\@@startlink{#1}\@@href}%
\providecommand \@@href[1]{\endgroup#1\@@endlink}%
\providecommand \@sanitize@url [0]{\catcode `\\12\catcode `\$12\catcode
  `\&12\catcode `\#12\catcode `\^12\catcode `\_12\catcode `\%12\relax}%
\providecommand \@@startlink[1]{}%
\providecommand \@@endlink[0]{}%
\providecommand \url  [0]{\begingroup\@sanitize@url \@url }%
\providecommand \@url [1]{\endgroup\@href {#1}{\urlprefix }}%
\providecommand \urlprefix  [0]{URL }%
\providecommand \Eprint [0]{\href }%
\providecommand \doibase [0]{http://dx.doi.org/}%
\providecommand \selectlanguage [0]{\@gobble}%
\providecommand \bibinfo  [0]{\@secondoftwo}%
\providecommand \bibfield  [0]{\@secondoftwo}%
\providecommand \translation [1]{[#1]}%
\providecommand \BibitemOpen [0]{}%
\providecommand \bibitemStop [0]{}%
\providecommand \bibitemNoStop [0]{.\EOS\space}%
\providecommand \EOS [0]{\spacefactor3000\relax}%
\providecommand \BibitemShut  [1]{\csname bibitem#1\endcsname}%
\let\auto@bib@innerbib\@empty
\bibitem [{\citenamefont {Spergel}\ and\ \citenamefont
  {Steinhardt}(2000)}]{Spergel:1999mh}%
  \BibitemOpen
  \bibfield  {author} {\bibinfo {author} {\bibfnamefont {D.~N.}\ \bibnamefont
  {Spergel}}\ and\ \bibinfo {author} {\bibfnamefont {P.~J.}\ \bibnamefont
  {Steinhardt}},\ }\href {\doibase 10.1103/PhysRevLett.84.3760} {\bibfield
  {journal} {\bibinfo  {journal} {Phys. Rev. Lett.}\ }\textbf {\bibinfo
  {volume} {84}},\ \bibinfo {pages} {3760} (\bibinfo {year} {2000})},\ \Eprint
  {http://arxiv.org/abs/astro-ph/9909386} {arXiv:astro-ph/9909386 [astro-ph]}
  \BibitemShut {NoStop}%
\bibitem [{\citenamefont {Firmani}\ \emph {et~al.}(2000)\citenamefont
  {Firmani}, \citenamefont {D'Onghia}, \citenamefont {Avila-Reese},
  \citenamefont {Chincarini},\ and\ \citenamefont
  {Hernandez}}]{Firmani:2000ce}%
  \BibitemOpen
  \bibfield  {author} {\bibinfo {author} {\bibfnamefont {C.}~\bibnamefont
  {Firmani}}, \bibinfo {author} {\bibfnamefont {E.}~\bibnamefont {D'Onghia}},
  \bibinfo {author} {\bibfnamefont {V.}~\bibnamefont {Avila-Reese}}, \bibinfo
  {author} {\bibfnamefont {G.}~\bibnamefont {Chincarini}}, \ and\ \bibinfo
  {author} {\bibfnamefont {X.}~\bibnamefont {Hernandez}},\ }\href {\doibase
  10.1046/j.1365-8711.2000.03555.x} {\bibfield  {journal} {\bibinfo  {journal}
  {Mon. Not. Roy. Astron. Soc.}\ }\textbf {\bibinfo {volume} {315}},\ \bibinfo
  {pages} {L29} (\bibinfo {year} {2000})},\ \Eprint
  {http://arxiv.org/abs/astro-ph/0002376} {arXiv:astro-ph/0002376 [astro-ph]}
  \BibitemShut {NoStop}%
\bibitem [{\citenamefont {Kaplinghat}\ \emph {et~al.}(2016)\citenamefont
  {Kaplinghat}, \citenamefont {Tulin},\ and\ \citenamefont
  {Yu}}]{Kaplinghat:2015aga}%
  \BibitemOpen
  \bibfield  {author} {\bibinfo {author} {\bibfnamefont {M.}~\bibnamefont
  {Kaplinghat}}, \bibinfo {author} {\bibfnamefont {S.}~\bibnamefont {Tulin}}, \
  and\ \bibinfo {author} {\bibfnamefont {H.-B.}\ \bibnamefont {Yu}},\ }\href
  {\doibase 10.1103/PhysRevLett.116.041302} {\bibfield  {journal} {\bibinfo
  {journal} {Phys. Rev. Lett.}\ }\textbf {\bibinfo {volume} {116}},\ \bibinfo
  {pages} {041302} (\bibinfo {year} {2016})},\ \Eprint
  {http://arxiv.org/abs/1508.03339} {arXiv:1508.03339 [astro-ph.CO]}
  \BibitemShut {NoStop}%
\bibitem [{\citenamefont {Bullock}\ and\ \citenamefont
  {Boylan-Kolchin}(2017)}]{Bullock:2017xww}%
  \BibitemOpen
  \bibfield  {author} {\bibinfo {author} {\bibfnamefont {J.~S.}\ \bibnamefont
  {Bullock}}\ and\ \bibinfo {author} {\bibfnamefont {M.}~\bibnamefont
  {Boylan-Kolchin}},\ }\href {\doibase 10.1146/annurev-astro-091916-055313}
  {\bibfield  {journal} {\bibinfo  {journal} {Ann. Rev. Astron. Astrophys.}\
  }\textbf {\bibinfo {volume} {55}},\ \bibinfo {pages} {343} (\bibinfo {year}
  {2017})},\ \Eprint {http://arxiv.org/abs/1707.04256} {arXiv:1707.04256
  [astro-ph.CO]} \BibitemShut {NoStop}%
\bibitem [{\citenamefont {Vogelsberger}\ \emph {et~al.}(2012)\citenamefont
  {Vogelsberger}, \citenamefont {Zavala},\ and\ \citenamefont
  {Loeb}}]{Vogelsberger:2012ku}%
  \BibitemOpen
  \bibfield  {author} {\bibinfo {author} {\bibfnamefont {M.}~\bibnamefont
  {Vogelsberger}}, \bibinfo {author} {\bibfnamefont {J.}~\bibnamefont
  {Zavala}}, \ and\ \bibinfo {author} {\bibfnamefont {A.}~\bibnamefont
  {Loeb}},\ }\href {\doibase 10.1111/j.1365-2966.2012.21182.x} {\bibfield
  {journal} {\bibinfo  {journal} {Mon. Not. Roy. Astron. Soc.}\ }\textbf
  {\bibinfo {volume} {423}},\ \bibinfo {pages} {3740} (\bibinfo {year}
  {2012})},\ \Eprint {http://arxiv.org/abs/1201.5892} {arXiv:1201.5892
  [astro-ph.CO]} \BibitemShut {NoStop}%
\bibitem [{\citenamefont {Rocha}\ \emph {et~al.}(2013)\citenamefont {Rocha},
  \citenamefont {Peter}, \citenamefont {Bullock}, \citenamefont {Kaplinghat},
  \citenamefont {Garrison-Kimmel}, \citenamefont {Onorbe},\ and\ \citenamefont
  {Moustakas}}]{Rocha:2012jg}%
  \BibitemOpen
  \bibfield  {author} {\bibinfo {author} {\bibfnamefont {M.}~\bibnamefont
  {Rocha}}, \bibinfo {author} {\bibfnamefont {A.~H.~G.}\ \bibnamefont {Peter}},
  \bibinfo {author} {\bibfnamefont {J.~S.}\ \bibnamefont {Bullock}}, \bibinfo
  {author} {\bibfnamefont {M.}~\bibnamefont {Kaplinghat}}, \bibinfo {author}
  {\bibfnamefont {S.}~\bibnamefont {Garrison-Kimmel}}, \bibinfo {author}
  {\bibfnamefont {J.}~\bibnamefont {Onorbe}}, \ and\ \bibinfo {author}
  {\bibfnamefont {L.~A.}\ \bibnamefont {Moustakas}},\ }\href {\doibase
  10.1093/mnras/sts514} {\bibfield  {journal} {\bibinfo  {journal} {Mon. Not.
  Roy. Astron. Soc.}\ }\textbf {\bibinfo {volume} {430}},\ \bibinfo {pages}
  {81} (\bibinfo {year} {2013})},\ \Eprint {http://arxiv.org/abs/1208.3025}
  {arXiv:1208.3025 [astro-ph.CO]} \BibitemShut {NoStop}%
\bibitem [{\citenamefont {Peter}\ \emph {et~al.}(2013)\citenamefont {Peter},
  \citenamefont {Rocha}, \citenamefont {Bullock},\ and\ \citenamefont
  {Kaplinghat}}]{Peter:2012jh}%
  \BibitemOpen
  \bibfield  {author} {\bibinfo {author} {\bibfnamefont {A.~H.~G.}\
  \bibnamefont {Peter}}, \bibinfo {author} {\bibfnamefont {M.}~\bibnamefont
  {Rocha}}, \bibinfo {author} {\bibfnamefont {J.~S.}\ \bibnamefont {Bullock}},
  \ and\ \bibinfo {author} {\bibfnamefont {M.}~\bibnamefont {Kaplinghat}},\
  }\href {\doibase 10.1093/mnras/sts535} {\bibfield  {journal} {\bibinfo
  {journal} {Mon. Not. Roy. Astron. Soc.}\ }\textbf {\bibinfo {volume} {430}},\
  \bibinfo {pages} {105} (\bibinfo {year} {2013})},\ \Eprint
  {http://arxiv.org/abs/1208.3026} {arXiv:1208.3026 [astro-ph.CO]} \BibitemShut
  {NoStop}%
\bibitem [{\citenamefont {Zavala}\ \emph {et~al.}(2013)\citenamefont {Zavala},
  \citenamefont {Vogelsberger},\ and\ \citenamefont {Walker}}]{Zavala:2012us}%
  \BibitemOpen
  \bibfield  {author} {\bibinfo {author} {\bibfnamefont {J.}~\bibnamefont
  {Zavala}}, \bibinfo {author} {\bibfnamefont {M.}~\bibnamefont
  {Vogelsberger}}, \ and\ \bibinfo {author} {\bibfnamefont {M.~G.}\
  \bibnamefont {Walker}},\ }\href {\doibase 10.1093/mnrasl/sls053} {\bibfield
  {journal} {\bibinfo  {journal} {Mon. Not. Roy. Astron. Soc.}\ }\textbf
  {\bibinfo {volume} {431}},\ \bibinfo {pages} {L20} (\bibinfo {year}
  {2013})},\ \Eprint {http://arxiv.org/abs/1211.6426} {arXiv:1211.6426
  [astro-ph.CO]} \BibitemShut {NoStop}%
\bibitem [{\citenamefont {Vogelsberger}\ \emph {et~al.}(2014)\citenamefont
  {Vogelsberger}, \citenamefont {Zavala}, \citenamefont {Simpson},\ and\
  \citenamefont {Jenkins}}]{Vogelsberger:2014pda}%
  \BibitemOpen
  \bibfield  {author} {\bibinfo {author} {\bibfnamefont {M.}~\bibnamefont
  {Vogelsberger}}, \bibinfo {author} {\bibfnamefont {J.}~\bibnamefont
  {Zavala}}, \bibinfo {author} {\bibfnamefont {C.}~\bibnamefont {Simpson}}, \
  and\ \bibinfo {author} {\bibfnamefont {A.}~\bibnamefont {Jenkins}},\ }\href
  {\doibase 10.1093/mnras/stu1713} {\bibfield  {journal} {\bibinfo  {journal}
  {Mon. Not. Roy. Astron. Soc.}\ }\textbf {\bibinfo {volume} {444}},\ \bibinfo
  {pages} {3684} (\bibinfo {year} {2014})},\ \Eprint
  {http://arxiv.org/abs/1405.5216} {arXiv:1405.5216 [astro-ph.CO]} \BibitemShut
  {NoStop}%
\bibitem [{\citenamefont {Fry}\ \emph {et~al.}(2015)\citenamefont {Fry},
  \citenamefont {Governato}, \citenamefont {Pontzen}, \citenamefont {Quinn},
  \citenamefont {Tremmel}, \citenamefont {Anderson}, \citenamefont {Menon},
  \citenamefont {Brooks},\ and\ \citenamefont {Wadsley}}]{Fry:2015rta}%
  \BibitemOpen
  \bibfield  {author} {\bibinfo {author} {\bibfnamefont {A.~B.}\ \bibnamefont
  {Fry}}, \bibinfo {author} {\bibfnamefont {F.}~\bibnamefont {Governato}},
  \bibinfo {author} {\bibfnamefont {A.}~\bibnamefont {Pontzen}}, \bibinfo
  {author} {\bibfnamefont {T.}~\bibnamefont {Quinn}}, \bibinfo {author}
  {\bibfnamefont {M.}~\bibnamefont {Tremmel}}, \bibinfo {author} {\bibfnamefont
  {L.}~\bibnamefont {Anderson}}, \bibinfo {author} {\bibfnamefont
  {H.}~\bibnamefont {Menon}}, \bibinfo {author} {\bibfnamefont {A.~M.}\
  \bibnamefont {Brooks}}, \ and\ \bibinfo {author} {\bibfnamefont
  {J.}~\bibnamefont {Wadsley}},\ }\href {\doibase 10.1093/mnras/stv1330}
  {\bibfield  {journal} {\bibinfo  {journal} {Mon. Not. Roy. Astron. Soc.}\
  }\textbf {\bibinfo {volume} {452}},\ \bibinfo {pages} {1468} (\bibinfo {year}
  {2015})},\ \Eprint {http://arxiv.org/abs/1501.00497} {arXiv:1501.00497
  [astro-ph.CO]} \BibitemShut {NoStop}%
\bibitem [{\citenamefont {Fitts}\ \emph {et~al.}(2018)\citenamefont {Fitts}
  \emph {et~al.}}]{Fitts:2018ycl}%
  \BibitemOpen
  \bibfield  {author} {\bibinfo {author} {\bibfnamefont {A.}~\bibnamefont
  {Fitts}} \emph {et~al.},\ }\href@noop {} {\  (\bibinfo {year} {2018})},\
  \Eprint {http://arxiv.org/abs/1811.11791} {arXiv:1811.11791 [astro-ph.GA]}
  \BibitemShut {NoStop}%
\bibitem [{\citenamefont {de~Blok}(2010)}]{deBlok:2009sp}%
  \BibitemOpen
  \bibfield  {author} {\bibinfo {author} {\bibfnamefont {W.~J.~G.}\
  \bibnamefont {de~Blok}},\ }\href {\doibase 10.1155/2010/789293} {\bibfield
  {journal} {\bibinfo  {journal} {Adv. Astron.}\ }\textbf {\bibinfo {volume}
  {2010}},\ \bibinfo {pages} {789293} (\bibinfo {year} {2010})},\ \Eprint
  {http://arxiv.org/abs/0910.3538} {arXiv:0910.3538 [astro-ph.CO]} \BibitemShut
  {NoStop}%
\bibitem [{\citenamefont {Salucci}\ \emph {et~al.}(2012)\citenamefont
  {Salucci}, \citenamefont {Wilkinson}, \citenamefont {Walker}, \citenamefont
  {Gilmore}, \citenamefont {Grebel}, \citenamefont {Koch}, \citenamefont
  {Martins},\ and\ \citenamefont {Wyse}}]{Salucci:2011ee}%
  \BibitemOpen
  \bibfield  {author} {\bibinfo {author} {\bibfnamefont {P.}~\bibnamefont
  {Salucci}}, \bibinfo {author} {\bibfnamefont {M.~I.}\ \bibnamefont
  {Wilkinson}}, \bibinfo {author} {\bibfnamefont {M.~G.}\ \bibnamefont
  {Walker}}, \bibinfo {author} {\bibfnamefont {G.~F.}\ \bibnamefont {Gilmore}},
  \bibinfo {author} {\bibfnamefont {E.~K.}\ \bibnamefont {Grebel}}, \bibinfo
  {author} {\bibfnamefont {A.}~\bibnamefont {Koch}}, \bibinfo {author}
  {\bibfnamefont {C.~F.}\ \bibnamefont {Martins}}, \ and\ \bibinfo {author}
  {\bibfnamefont {R.~F.~G.}\ \bibnamefont {Wyse}},\ }\href {\doibase
  10.1111/j.1365-2966.2011.20144.x} {\bibfield  {journal} {\bibinfo  {journal}
  {Mon. Not. Roy. Astron. Soc.}\ }\textbf {\bibinfo {volume} {420}},\ \bibinfo
  {pages} {2034} (\bibinfo {year} {2012})},\ \Eprint
  {http://arxiv.org/abs/1111.1165} {arXiv:1111.1165 [astro-ph.CO]} \BibitemShut
  {NoStop}%
\bibitem [{\citenamefont {Kuzio~de Naray}\ \emph {et~al.}(2010)\citenamefont
  {Kuzio~de Naray}, \citenamefont {Martinez}, \citenamefont {Bullock},\ and\
  \citenamefont {Kaplinghat}}]{deNaray:2009xj}%
  \BibitemOpen
  \bibfield  {author} {\bibinfo {author} {\bibfnamefont {R.}~\bibnamefont
  {Kuzio~de Naray}}, \bibinfo {author} {\bibfnamefont {G.~D.}\ \bibnamefont
  {Martinez}}, \bibinfo {author} {\bibfnamefont {J.~S.}\ \bibnamefont
  {Bullock}}, \ and\ \bibinfo {author} {\bibfnamefont {M.}~\bibnamefont
  {Kaplinghat}},\ }\href {\doibase 10.1088/2041-8205/710/2/L161} {\bibfield
  {journal} {\bibinfo  {journal} {Astrophys. J.}\ }\textbf {\bibinfo {volume}
  {710}},\ \bibinfo {pages} {L161} (\bibinfo {year} {2010})},\ \Eprint
  {http://arxiv.org/abs/0912.3518} {arXiv:0912.3518 [astro-ph.CO]} \BibitemShut
  {NoStop}%
\bibitem [{\citenamefont {Oman}\ \emph {et~al.}(2015)\citenamefont {Oman} \emph
  {et~al.}}]{Oman:2015xda}%
  \BibitemOpen
  \bibfield  {author} {\bibinfo {author} {\bibfnamefont {K.~A.}\ \bibnamefont
  {Oman}} \emph {et~al.},\ }\href {\doibase 10.1093/mnras/stv1504} {\bibfield
  {journal} {\bibinfo  {journal} {Mon. Not. Roy. Astron. Soc.}\ }\textbf
  {\bibinfo {volume} {452}},\ \bibinfo {pages} {3650} (\bibinfo {year}
  {2015})},\ \Eprint {http://arxiv.org/abs/1504.01437} {arXiv:1504.01437
  [astro-ph.GA]} \BibitemShut {NoStop}%
\bibitem [{\citenamefont {McGaugh}\ \emph {et~al.}(2016)\citenamefont
  {McGaugh}, \citenamefont {Lelli},\ and\ \citenamefont
  {Schombert}}]{McGaugh:2016leg}%
  \BibitemOpen
  \bibfield  {author} {\bibinfo {author} {\bibfnamefont {S.}~\bibnamefont
  {McGaugh}}, \bibinfo {author} {\bibfnamefont {F.}~\bibnamefont {Lelli}}, \
  and\ \bibinfo {author} {\bibfnamefont {J.}~\bibnamefont {Schombert}},\ }\href
  {\doibase 10.1103/PhysRevLett.117.201101} {\bibfield  {journal} {\bibinfo
  {journal} {Phys. Rev. Lett.}\ }\textbf {\bibinfo {volume} {117}},\ \bibinfo
  {pages} {201101} (\bibinfo {year} {2016})},\ \Eprint
  {http://arxiv.org/abs/1609.05917} {arXiv:1609.05917 [astro-ph.GA]}
  \BibitemShut {NoStop}%
\bibitem [{\citenamefont {Kamada}\ \emph {et~al.}(2017)\citenamefont {Kamada},
  \citenamefont {Kaplinghat}, \citenamefont {Pace},\ and\ \citenamefont
  {Yu}}]{Kamada:2016euw}%
  \BibitemOpen
  \bibfield  {author} {\bibinfo {author} {\bibfnamefont {A.}~\bibnamefont
  {Kamada}}, \bibinfo {author} {\bibfnamefont {M.}~\bibnamefont {Kaplinghat}},
  \bibinfo {author} {\bibfnamefont {A.~B.}\ \bibnamefont {Pace}}, \ and\
  \bibinfo {author} {\bibfnamefont {H.-B.}\ \bibnamefont {Yu}},\ }\href
  {\doibase 10.1103/PhysRevLett.119.111102} {\bibfield  {journal} {\bibinfo
  {journal} {Phys. Rev. Lett.}\ }\textbf {\bibinfo {volume} {119}},\ \bibinfo
  {pages} {111102} (\bibinfo {year} {2017})},\ \Eprint
  {http://arxiv.org/abs/1611.02716} {arXiv:1611.02716 [astro-ph.GA]}
  \BibitemShut {NoStop}%
\bibitem [{\citenamefont {Ren}\ \emph {et~al.}(2018)\citenamefont {Ren},
  \citenamefont {Kwa}, \citenamefont {Kaplinghat},\ and\ \citenamefont
  {Hai-Bo}}]{Ren:2018jpt}%
  \BibitemOpen
  \bibfield  {author} {\bibinfo {author} {\bibfnamefont {T.}~\bibnamefont
  {Ren}}, \bibinfo {author} {\bibfnamefont {A.}~\bibnamefont {Kwa}}, \bibinfo
  {author} {\bibfnamefont {M.}~\bibnamefont {Kaplinghat}}, \ and\ \bibinfo
  {author} {\bibfnamefont {Y.}~\bibnamefont {Hai-Bo}},\ }\href@noop {} {\
  (\bibinfo {year} {2018})},\ \Eprint {http://arxiv.org/abs/1808.05695}
  {arXiv:1808.05695 [astro-ph.GA]} \BibitemShut {NoStop}%
\bibitem [{\citenamefont {Lynden-Bell}\ and\ \citenamefont
  {Eggleton}(1980)}]{doi:10.1093/mnras/191.3.483}%
  \BibitemOpen
  \bibfield  {author} {\bibinfo {author} {\bibfnamefont {D.}~\bibnamefont
  {Lynden-Bell}}\ and\ \bibinfo {author} {\bibfnamefont {P.~P.}\ \bibnamefont
  {Eggleton}},\ }\href {\doibase 10.1093/mnras/191.3.483} {\bibfield  {journal}
  {\bibinfo  {journal} {Mon. Not. Roy. Astron. Soc.}\ }\textbf {\bibinfo
  {volume} {191}},\ \bibinfo {pages} {483} (\bibinfo {year}
  {1980})}\BibitemShut {NoStop}%
\bibitem [{\citenamefont {Quinlan}(1996)}]{Quinlan:1996bw}%
  \BibitemOpen
  \bibfield  {author} {\bibinfo {author} {\bibfnamefont {G.~D.}\ \bibnamefont
  {Quinlan}},\ }\href {\doibase 10.1016/S1384-1076(96)00018-8} {\bibfield
  {journal} {\bibinfo  {journal} {New Astron.}\ }\textbf {\bibinfo {volume}
  {1}},\ \bibinfo {pages} {255} (\bibinfo {year} {1996})},\ \Eprint
  {http://arxiv.org/abs/astro-ph/9606182} {arXiv:astro-ph/9606182 [astro-ph]}
  \BibitemShut {NoStop}%
\bibitem [{\citenamefont {Balberg}\ \emph {et~al.}(2002)\citenamefont
  {Balberg}, \citenamefont {Shapiro},\ and\ \citenamefont
  {Inagaki}}]{Balberg:2002ue}%
  \BibitemOpen
  \bibfield  {author} {\bibinfo {author} {\bibfnamefont {S.}~\bibnamefont
  {Balberg}}, \bibinfo {author} {\bibfnamefont {S.~L.}\ \bibnamefont
  {Shapiro}}, \ and\ \bibinfo {author} {\bibfnamefont {S.}~\bibnamefont
  {Inagaki}},\ }\href {\doibase 10.1086/339038} {\bibfield  {journal} {\bibinfo
   {journal} {Astrophys. J.}\ }\textbf {\bibinfo {volume} {568}},\ \bibinfo
  {pages} {475} (\bibinfo {year} {2002})},\ \Eprint
  {http://arxiv.org/abs/astro-ph/0110561} {arXiv:astro-ph/0110561 [astro-ph]}
  \BibitemShut {NoStop}%
\bibitem [{\citenamefont {Koda}\ and\ \citenamefont
  {Shapiro}(2011)}]{Koda:2011yb}%
  \BibitemOpen
  \bibfield  {author} {\bibinfo {author} {\bibfnamefont {J.}~\bibnamefont
  {Koda}}\ and\ \bibinfo {author} {\bibfnamefont {P.~R.}\ \bibnamefont
  {Shapiro}},\ }\href {\doibase 10.1111/j.1365-2966.2011.18684.x} {\bibfield
  {journal} {\bibinfo  {journal} {Mon. Not. Roy. Astron. Soc.}\ }\textbf
  {\bibinfo {volume} {415}},\ \bibinfo {pages} {1125} (\bibinfo {year}
  {2011})},\ \Eprint {http://arxiv.org/abs/1101.3097} {arXiv:1101.3097
  [astro-ph.CO]} \BibitemShut {NoStop}%
\bibitem [{\citenamefont {Elbert}\ \emph {et~al.}(2015)\citenamefont {Elbert},
  \citenamefont {Bullock}, \citenamefont {Garrison-Kimmel}, \citenamefont
  {Rocha}, \citenamefont {Oñorbe},\ and\ \citenamefont
  {Peter}}]{Elbert:2014bma}%
  \BibitemOpen
  \bibfield  {author} {\bibinfo {author} {\bibfnamefont {O.~D.}\ \bibnamefont
  {Elbert}}, \bibinfo {author} {\bibfnamefont {J.~S.}\ \bibnamefont {Bullock}},
  \bibinfo {author} {\bibfnamefont {S.}~\bibnamefont {Garrison-Kimmel}},
  \bibinfo {author} {\bibfnamefont {M.}~\bibnamefont {Rocha}}, \bibinfo
  {author} {\bibfnamefont {J.}~\bibnamefont {Oñorbe}}, \ and\ \bibinfo
  {author} {\bibfnamefont {A.~H.~G.}\ \bibnamefont {Peter}},\ }\href {\doibase
  10.1093/mnras/stv1470} {\bibfield  {journal} {\bibinfo  {journal} {Mon. Not.
  Roy. Astron. Soc.}\ }\textbf {\bibinfo {volume} {453}},\ \bibinfo {pages}
  {29} (\bibinfo {year} {2015})},\ \Eprint {http://arxiv.org/abs/1412.1477}
  {arXiv:1412.1477 [astro-ph.GA]} \BibitemShut {NoStop}%
\bibitem [{\citenamefont {Chapman}\ \emph {et~al.}(1990)\citenamefont
  {Chapman}, \citenamefont {Cowling}, \citenamefont {Burnett},\ and\
  \citenamefont {Cercignani}}]{chapman1990mathematical}%
  \BibitemOpen
  \bibfield  {author} {\bibinfo {author} {\bibfnamefont {S.}~\bibnamefont
  {Chapman}}, \bibinfo {author} {\bibfnamefont {T.}~\bibnamefont {Cowling}},
  \bibinfo {author} {\bibfnamefont {D.}~\bibnamefont {Burnett}}, \ and\
  \bibinfo {author} {\bibfnamefont {C.}~\bibnamefont {Cercignani}},\ }\href
  {https://books.google.com/books?id=Cbp5JP2OTrwC} {\emph {\bibinfo {title}
  {The Mathematical Theory of Non-uniform Gases: An Account of the Kinetic
  Theory of Viscosity, Thermal Conduction and Diffusion in Gases}}},\ Cambridge
  Mathematical Library\ (\bibinfo  {publisher} {Cambridge University Press},\
  \bibinfo {year} {1990})\BibitemShut {NoStop}%
\bibitem [{\citenamefont {Pollack}\ \emph {et~al.}(2015)\citenamefont
  {Pollack}, \citenamefont {Spergel},\ and\ \citenamefont
  {Steinhardt}}]{Pollack:2014rja}%
  \BibitemOpen
  \bibfield  {author} {\bibinfo {author} {\bibfnamefont {J.}~\bibnamefont
  {Pollack}}, \bibinfo {author} {\bibfnamefont {D.~N.}\ \bibnamefont
  {Spergel}}, \ and\ \bibinfo {author} {\bibfnamefont {P.~J.}\ \bibnamefont
  {Steinhardt}},\ }\href {\doibase 10.1088/0004-637X/804/2/131} {\bibfield
  {journal} {\bibinfo  {journal} {Astrophys. J.}\ }\textbf {\bibinfo {volume}
  {804}},\ \bibinfo {pages} {131} (\bibinfo {year} {2015})},\ \Eprint
  {http://arxiv.org/abs/1501.00017} {arXiv:1501.00017 [astro-ph.CO]}
  \BibitemShut {NoStop}%
\bibitem [{\citenamefont {Li}\ \emph {et~al.}(2018)\citenamefont {Li} \emph
  {et~al.}}]{Li:2018zfs}%
  \BibitemOpen
  \bibfield  {author} {\bibinfo {author} {\bibfnamefont {T.~S.}\ \bibnamefont
  {Li}} \emph {et~al.} (\bibinfo {collaboration} {DES}),\ }\href {\doibase
  10.3847/1538-4357/aadf91} {\bibfield  {journal} {\bibinfo  {journal}
  {Astrophys. J.}\ }\textbf {\bibinfo {volume} {866}},\ \bibinfo {pages} {22}
  (\bibinfo {year} {2018})},\ \Eprint {http://arxiv.org/abs/1804.07761}
  {arXiv:1804.07761 [astro-ph.GA]} \BibitemShut {NoStop}%
\bibitem [{\citenamefont {Simon}\ \emph {et~al.}(2017)\citenamefont {Simon}
  \emph {et~al.}}]{Simon:2016mkr}%
  \BibitemOpen
  \bibfield  {author} {\bibinfo {author} {\bibfnamefont {J.~D.}\ \bibnamefont
  {Simon}} \emph {et~al.} (\bibinfo {collaboration} {DES}),\ }\href {\doibase
  10.3847/1538-4357/aa5be7} {\bibfield  {journal} {\bibinfo  {journal}
  {Astrophys. J.}\ }\textbf {\bibinfo {volume} {838}},\ \bibinfo {pages} {11}
  (\bibinfo {year} {2017})},\ \Eprint {http://arxiv.org/abs/1610.05301}
  {arXiv:1610.05301 [astro-ph.GA]} \BibitemShut {NoStop}%
\bibitem [{\citenamefont {Kirby}\ \emph {et~al.}(2015)\citenamefont {Kirby},
  \citenamefont {Cohen}, \citenamefont {Simon},\ and\ \citenamefont
  {Guhathakurta}}]{Kirby:2015bxa}%
  \BibitemOpen
  \bibfield  {author} {\bibinfo {author} {\bibfnamefont {E.~N.}\ \bibnamefont
  {Kirby}}, \bibinfo {author} {\bibfnamefont {J.~G.}\ \bibnamefont {Cohen}},
  \bibinfo {author} {\bibfnamefont {J.~D.}\ \bibnamefont {Simon}}, \ and\
  \bibinfo {author} {\bibfnamefont {P.}~\bibnamefont {Guhathakurta}},\ }\href
  {\doibase 10.1088/2041-8205/814/1/L7} {\bibfield  {journal} {\bibinfo
  {journal} {Astrophys. J.}\ }\textbf {\bibinfo {volume} {814}},\ \bibinfo
  {pages} {L7} (\bibinfo {year} {2015})},\ \Eprint
  {http://arxiv.org/abs/1510.03856} {arXiv:1510.03856 [astro-ph.GA]}
  \BibitemShut {NoStop}%
\bibitem [{\citenamefont {Teyssier}\ \emph {et~al.}(2012)\citenamefont
  {Teyssier}, \citenamefont {Johnston},\ and\ \citenamefont
  {Kuhlen}}]{Teyssier:2012ay}%
  \BibitemOpen
  \bibfield  {author} {\bibinfo {author} {\bibfnamefont {M.}~\bibnamefont
  {Teyssier}}, \bibinfo {author} {\bibfnamefont {K.~V.}\ \bibnamefont
  {Johnston}}, \ and\ \bibinfo {author} {\bibfnamefont {M.}~\bibnamefont
  {Kuhlen}},\ }\href {\doibase 10.1111/j.1365-2966.2012.21793.x} {\bibfield
  {journal} {\bibinfo  {journal} {Mon. Not. Roy. Astron. Soc.}\ }\textbf
  {\bibinfo {volume} {426}},\ \bibinfo {pages} {1808} (\bibinfo {year}
  {2012})},\ \Eprint {http://arxiv.org/abs/1207.2768} {arXiv:1207.2768
  [astro-ph.GA]} \BibitemShut {NoStop}%
\bibitem [{\citenamefont {Penarrubia}\ \emph {et~al.}(2010)\citenamefont
  {Penarrubia}, \citenamefont {Benson}, \citenamefont {Walker}, \citenamefont
  {Gilmore}, \citenamefont {McConnachie},\ and\ \citenamefont
  {Mayer}}]{Penarrubia:2010jk}%
  \BibitemOpen
  \bibfield  {author} {\bibinfo {author} {\bibfnamefont {J.}~\bibnamefont
  {Penarrubia}}, \bibinfo {author} {\bibfnamefont {A.~J.}\ \bibnamefont
  {Benson}}, \bibinfo {author} {\bibfnamefont {M.~G.}\ \bibnamefont {Walker}},
  \bibinfo {author} {\bibfnamefont {G.}~\bibnamefont {Gilmore}}, \bibinfo
  {author} {\bibfnamefont {A.}~\bibnamefont {McConnachie}}, \ and\ \bibinfo
  {author} {\bibfnamefont {L.}~\bibnamefont {Mayer}},\ }\href {\doibase
  10.1111/j.1365-2966.2010.16762.x} {\bibfield  {journal} {\bibinfo  {journal}
  {Mon. Not. Roy. Astron. Soc.}\ }\textbf {\bibinfo {volume} {406}},\ \bibinfo
  {pages} {1290} (\bibinfo {year} {2010})},\ \Eprint
  {http://arxiv.org/abs/1002.3376} {arXiv:1002.3376 [astro-ph.GA]} \BibitemShut
  {NoStop}%
\bibitem [{\citenamefont {Diemer}(2018)}]{Diemer:2017bwl}%
  \BibitemOpen
  \bibfield  {author} {\bibinfo {author} {\bibfnamefont {B.}~\bibnamefont
  {Diemer}},\ }\href {\doibase 10.3847/1538-4365/aaee8c} {\bibfield  {journal}
  {\bibinfo  {journal} {Astrophys. J. Suppl.}\ }\textbf {\bibinfo {volume}
  {239}},\ \bibinfo {pages} {35} (\bibinfo {year} {2018})},\ \Eprint
  {http://arxiv.org/abs/1712.04512} {arXiv:1712.04512 [astro-ph.CO]}
  \BibitemShut {NoStop}%
\bibitem [{\citenamefont {Rocha}\ \emph {et~al.}(2012)\citenamefont {Rocha},
  \citenamefont {Peter},\ and\ \citenamefont {Bullock}}]{Rocha:2011aa}%
  \BibitemOpen
  \bibfield  {author} {\bibinfo {author} {\bibfnamefont {M.}~\bibnamefont
  {Rocha}}, \bibinfo {author} {\bibfnamefont {A.~H.~G.}\ \bibnamefont {Peter}},
  \ and\ \bibinfo {author} {\bibfnamefont {J.~S.}\ \bibnamefont {Bullock}},\
  }\href {\doibase 10.1111/j.1365-2966.2012.21432.x} {\bibfield  {journal}
  {\bibinfo  {journal} {Mon. Not. Roy. Astron. Soc.}\ }\textbf {\bibinfo
  {volume} {425}},\ \bibinfo {pages} {231} (\bibinfo {year} {2012})},\ \Eprint
  {http://arxiv.org/abs/1110.0464} {arXiv:1110.0464 [astro-ph.CO]} \BibitemShut
  {NoStop}%
\bibitem [{\citenamefont {Garrison-Kimmel}\ \emph {et~al.}(2014)\citenamefont
  {Garrison-Kimmel}, \citenamefont {Boylan-Kolchin}, \citenamefont {Bullock},\
  and\ \citenamefont {Kirby}}]{Garrison-Kimmel:2014vqa}%
  \BibitemOpen
  \bibfield  {author} {\bibinfo {author} {\bibfnamefont {S.}~\bibnamefont
  {Garrison-Kimmel}}, \bibinfo {author} {\bibfnamefont {M.}~\bibnamefont
  {Boylan-Kolchin}}, \bibinfo {author} {\bibfnamefont {J.~S.}\ \bibnamefont
  {Bullock}}, \ and\ \bibinfo {author} {\bibfnamefont {E.~N.}\ \bibnamefont
  {Kirby}},\ }\href {\doibase 10.1093/mnras/stu1477} {\bibfield  {journal}
  {\bibinfo  {journal} {Mon. Not. Roy. Astron. Soc.}\ }\textbf {\bibinfo
  {volume} {444}},\ \bibinfo {pages} {222} (\bibinfo {year} {2014})},\ \Eprint
  {http://arxiv.org/abs/1404.5313} {arXiv:1404.5313 [astro-ph.GA]} \BibitemShut
  {NoStop}%
\bibitem [{\citenamefont {Boylan-Kolchin}\ \emph {et~al.}(2012)\citenamefont
  {Boylan-Kolchin}, \citenamefont {Bullock},\ and\ \citenamefont
  {Kaplinghat}}]{BoylanKolchin:2011dk}%
  \BibitemOpen
  \bibfield  {author} {\bibinfo {author} {\bibfnamefont {M.}~\bibnamefont
  {Boylan-Kolchin}}, \bibinfo {author} {\bibfnamefont {J.~S.}\ \bibnamefont
  {Bullock}}, \ and\ \bibinfo {author} {\bibfnamefont {M.}~\bibnamefont
  {Kaplinghat}},\ }\href {\doibase 10.1111/j.1365-2966.2012.20695.x} {\bibfield
   {journal} {\bibinfo  {journal} {Mon. Not. Roy. Astron. Soc.}\ }\textbf
  {\bibinfo {volume} {422}},\ \bibinfo {pages} {1203} (\bibinfo {year}
  {2012})},\ \Eprint {http://arxiv.org/abs/1111.2048} {arXiv:1111.2048
  [astro-ph.CO]} \BibitemShut {NoStop}%
\bibitem [{\citenamefont {Fraternali}\ \emph {et~al.}(2009)\citenamefont
  {Fraternali}, \citenamefont {Tolstoy}, \citenamefont {Irwin},\ and\
  \citenamefont {Cole}}]{Fraternali:2009de}%
  \BibitemOpen
  \bibfield  {author} {\bibinfo {author} {\bibfnamefont {F.}~\bibnamefont
  {Fraternali}}, \bibinfo {author} {\bibfnamefont {E.}~\bibnamefont {Tolstoy}},
  \bibinfo {author} {\bibfnamefont {M.}~\bibnamefont {Irwin}}, \ and\ \bibinfo
  {author} {\bibfnamefont {A.}~\bibnamefont {Cole}},\ }\href {\doibase
  10.1051/0004-6361/200810830} {\bibfield  {journal} {\bibinfo  {journal}
  {Astron. Astrophys.}\ }\textbf {\bibinfo {volume} {499}},\ \bibinfo {pages}
  {121} (\bibinfo {year} {2009})},\ \Eprint {http://arxiv.org/abs/0903.4635}
  {arXiv:0903.4635 [astro-ph.CO]} \BibitemShut {NoStop}%
\bibitem [{\citenamefont {Kirby}\ \emph {et~al.}(2014)\citenamefont {Kirby},
  \citenamefont {Bullock}, \citenamefont {Boylan-Kolchin}, \citenamefont
  {Kaplinghat},\ and\ \citenamefont {Cohen}}]{Kirby:2014sya}%
  \BibitemOpen
  \bibfield  {author} {\bibinfo {author} {\bibfnamefont {E.~N.}\ \bibnamefont
  {Kirby}}, \bibinfo {author} {\bibfnamefont {J.~S.}\ \bibnamefont {Bullock}},
  \bibinfo {author} {\bibfnamefont {M.}~\bibnamefont {Boylan-Kolchin}},
  \bibinfo {author} {\bibfnamefont {M.}~\bibnamefont {Kaplinghat}}, \ and\
  \bibinfo {author} {\bibfnamefont {J.~G.}\ \bibnamefont {Cohen}},\ }\href
  {\doibase 10.1093/mnras/stu025} {\bibfield  {journal} {\bibinfo  {journal}
  {Mon. Not. Roy. Astron. Soc.}\ }\textbf {\bibinfo {volume} {439}},\ \bibinfo
  {pages} {1015} (\bibinfo {year} {2014})},\ \Eprint
  {http://arxiv.org/abs/1401.1208} {arXiv:1401.1208 [astro-ph.GA]} \BibitemShut
  {NoStop}%
\bibitem [{\citenamefont {Sales}\ \emph {et~al.}(2007)\citenamefont {Sales},
  \citenamefont {Navarro}, \citenamefont {Abadi},\ and\ \citenamefont
  {Steinmetz}}]{Sales:2007hr}%
  \BibitemOpen
  \bibfield  {author} {\bibinfo {author} {\bibfnamefont {L.~V.}\ \bibnamefont
  {Sales}}, \bibinfo {author} {\bibfnamefont {J.~F.}\ \bibnamefont {Navarro}},
  \bibinfo {author} {\bibfnamefont {M.~G.}\ \bibnamefont {Abadi}}, \ and\
  \bibinfo {author} {\bibfnamefont {M.}~\bibnamefont {Steinmetz}},\ }\href
  {\doibase 10.1111/j.1365-2966.2007.12026.x} {\bibfield  {journal} {\bibinfo
  {journal} {Mon. Not. Roy. Astron. Soc.}\ }\textbf {\bibinfo {volume} {379}},\
  \bibinfo {pages} {1475} (\bibinfo {year} {2007})},\ \Eprint
  {http://arxiv.org/abs/0704.1773} {arXiv:0704.1773 [astro-ph]} \BibitemShut
  {NoStop}%
\bibitem [{\citenamefont {Saviane}\ and\ \citenamefont
  {Held}(1996)}]{Saviane:1996xf}%
  \BibitemOpen
  \bibfield  {author} {\bibinfo {author} {\bibfnamefont {I.}~\bibnamefont
  {Saviane}}\ and\ \bibinfo {author} {\bibfnamefont {E.~V.}\ \bibnamefont
  {Held}},\ }\href@noop {} {\bibfield  {journal} {\bibinfo  {journal} {Astron.
  Astrophys.}\ }\textbf {\bibinfo {volume} {315}},\ \bibinfo {pages} {40}
  (\bibinfo {year} {1996})},\ \Eprint {http://arxiv.org/abs/astro-ph/9601165}
  {arXiv:astro-ph/9601165 [astro-ph]} \BibitemShut {NoStop}%
\bibitem [{\citenamefont {{Monelli}}\ \emph {et~al.}(2010)\citenamefont
  {{Monelli}}, \citenamefont {{Gallart}}, \citenamefont {{Hidalgo}},
  \citenamefont {{Aparicio}}, \citenamefont {{Skillman}}, \citenamefont
  {{Cole}}, \citenamefont {{Weisz}}, \citenamefont {{Mayer}}, \citenamefont
  {{Bernard}}, \citenamefont {{Cassisi}}, \citenamefont {{Dolphin}},
  \citenamefont {{Drozdovsky}},\ and\ \citenamefont
  {{Stetson}}}]{2010ApJ...722.1864M}%
  \BibitemOpen
  \bibfield  {author} {\bibinfo {author} {\bibfnamefont {M.}~\bibnamefont
  {{Monelli}}}, \bibinfo {author} {\bibfnamefont {C.}~\bibnamefont
  {{Gallart}}}, \bibinfo {author} {\bibfnamefont {S.~L.}\ \bibnamefont
  {{Hidalgo}}}, \bibinfo {author} {\bibfnamefont {A.}~\bibnamefont
  {{Aparicio}}}, \bibinfo {author} {\bibfnamefont {E.~D.}\ \bibnamefont
  {{Skillman}}}, \bibinfo {author} {\bibfnamefont {A.~A.}\ \bibnamefont
  {{Cole}}}, \bibinfo {author} {\bibfnamefont {D.~R.}\ \bibnamefont {{Weisz}}},
  \bibinfo {author} {\bibfnamefont {L.}~\bibnamefont {{Mayer}}}, \bibinfo
  {author} {\bibfnamefont {E.~J.}\ \bibnamefont {{Bernard}}}, \bibinfo {author}
  {\bibfnamefont {S.}~\bibnamefont {{Cassisi}}}, \bibinfo {author}
  {\bibfnamefont {A.~E.}\ \bibnamefont {{Dolphin}}}, \bibinfo {author}
  {\bibfnamefont {I.}~\bibnamefont {{Drozdovsky}}}, \ and\ \bibinfo {author}
  {\bibfnamefont {P.~B.}\ \bibnamefont {{Stetson}}},\ }\href {\doibase
  10.1088/0004-637X/722/2/1864} {\bibfield  {journal} {\bibinfo  {journal}
  {Astrophys. J.}\ }\textbf {\bibinfo {volume} {722}},\ \bibinfo {pages} {1864}
  (\bibinfo {year} {2010})},\ \Eprint {http://arxiv.org/abs/1010.2982}
  {arXiv:1010.2982} \BibitemShut {NoStop}%
\bibitem [{\citenamefont {Read}\ \emph {et~al.}(2018)\citenamefont {Read},
  \citenamefont {Walker},\ and\ \citenamefont {Steger}}]{Read:2018pft}%
  \BibitemOpen
  \bibfield  {author} {\bibinfo {author} {\bibfnamefont {J.~I.}\ \bibnamefont
  {Read}}, \bibinfo {author} {\bibfnamefont {M.~G.}\ \bibnamefont {Walker}}, \
  and\ \bibinfo {author} {\bibfnamefont {P.}~\bibnamefont {Steger}},\ }\href
  {\doibase 10.1093/mnras/sty2286} {\  (\bibinfo {year} {2018}),\
  10.1093/mnras/sty2286},\ \Eprint {http://arxiv.org/abs/1805.06934}
  {arXiv:1805.06934 [astro-ph.GA]} \BibitemShut {NoStop}%
\bibitem [{\citenamefont {Valli}\ and\ \citenamefont
  {Yu}(2018)}]{Valli:2017ktb}%
  \BibitemOpen
  \bibfield  {author} {\bibinfo {author} {\bibfnamefont {M.}~\bibnamefont
  {Valli}}\ and\ \bibinfo {author} {\bibfnamefont {H.-B.}\ \bibnamefont {Yu}},\
  }\href {\doibase 10.1038/s41550-018-0560-7} {\bibfield  {journal} {\bibinfo
  {journal} {Nat. Astron.}\ }\textbf {\bibinfo {volume} {2}},\ \bibinfo {pages}
  {907} (\bibinfo {year} {2018})},\ \Eprint {http://arxiv.org/abs/1711.03502}
  {arXiv:1711.03502 [astro-ph.GA]} \BibitemShut {NoStop}%
\bibitem [{\citenamefont {{Fritz}}\ \emph {et~al.}(2018)\citenamefont
  {{Fritz}}, \citenamefont {{Battaglia}}, \citenamefont {{Pawlowski}},
  \citenamefont {{Kallivayalil}}, \citenamefont {{van der Marel}},
  \citenamefont {{Sohn}}, \citenamefont {{Brook}},\ and\ \citenamefont
  {{Besla}}}]{2018A&A...619A.103F}%
  \BibitemOpen
  \bibfield  {author} {\bibinfo {author} {\bibfnamefont {T.~K.}\ \bibnamefont
  {{Fritz}}}, \bibinfo {author} {\bibfnamefont {G.}~\bibnamefont
  {{Battaglia}}}, \bibinfo {author} {\bibfnamefont {M.~S.}\ \bibnamefont
  {{Pawlowski}}}, \bibinfo {author} {\bibfnamefont {N.}~\bibnamefont
  {{Kallivayalil}}}, \bibinfo {author} {\bibfnamefont {R.}~\bibnamefont {{van
  der Marel}}}, \bibinfo {author} {\bibfnamefont {S.~T.}\ \bibnamefont
  {{Sohn}}}, \bibinfo {author} {\bibfnamefont {C.}~\bibnamefont {{Brook}}}, \
  and\ \bibinfo {author} {\bibfnamefont {G.}~\bibnamefont {{Besla}}},\ }\href
  {\doibase 10.1051/0004-6361/201833343} {\bibfield  {journal} {\bibinfo
  {journal} {Astron. Astrophys.}\ }\textbf {\bibinfo {volume} {619}},\ \bibinfo
  {eid} {A103} (\bibinfo {year} {2018})},\ \Eprint
  {http://arxiv.org/abs/1805.00908} {arXiv:1805.00908} \BibitemShut {NoStop}%
\bibitem [{\citenamefont {D'Onghia}\ \emph {et~al.}(2010)\citenamefont
  {D'Onghia}, \citenamefont {Springel}, \citenamefont {Hernquist},\ and\
  \citenamefont {Keres}}]{DOnghia:2009xhq}%
  \BibitemOpen
  \bibfield  {author} {\bibinfo {author} {\bibfnamefont {E.}~\bibnamefont
  {D'Onghia}}, \bibinfo {author} {\bibfnamefont {V.}~\bibnamefont {Springel}},
  \bibinfo {author} {\bibfnamefont {L.}~\bibnamefont {Hernquist}}, \ and\
  \bibinfo {author} {\bibfnamefont {D.}~\bibnamefont {Keres}},\ }\href
  {\doibase 10.1088/0004-637X/709/2/1138} {\bibfield  {journal} {\bibinfo
  {journal} {Astrophys. J.}\ }\textbf {\bibinfo {volume} {709}},\ \bibinfo
  {pages} {1138} (\bibinfo {year} {2010})},\ \Eprint
  {http://arxiv.org/abs/0907.3482} {arXiv:0907.3482 [astro-ph.CO]} \BibitemShut
  {NoStop}%
\bibitem [{\citenamefont {Garrison-Kimmel}\ \emph {et~al.}(2017)\citenamefont
  {Garrison-Kimmel} \emph {et~al.}}]{Garrison-Kimmel:2017zes}%
  \BibitemOpen
  \bibfield  {author} {\bibinfo {author} {\bibfnamefont {S.}~\bibnamefont
  {Garrison-Kimmel}} \emph {et~al.},\ }\href {\doibase 10.1093/mnras/stx1710}
  {\bibfield  {journal} {\bibinfo  {journal} {Mon. Not. Roy. Astron. Soc.}\
  }\textbf {\bibinfo {volume} {471}},\ \bibinfo {pages} {1709} (\bibinfo {year}
  {2017})},\ \Eprint {http://arxiv.org/abs/1701.03792} {arXiv:1701.03792
  [astro-ph.GA]} \BibitemShut {NoStop}%
\bibitem [{\citenamefont {Kelley}\ \emph {et~al.}(2018)\citenamefont {Kelley},
  \citenamefont {Bullock}, \citenamefont {Garrison-Kimmel}, \citenamefont
  {Boylan-Kolchin}, \citenamefont {Pawlowski},\ and\ \citenamefont
  {Graus}}]{Kelley:2018pdy}%
  \BibitemOpen
  \bibfield  {author} {\bibinfo {author} {\bibfnamefont {T.}~\bibnamefont
  {Kelley}}, \bibinfo {author} {\bibfnamefont {J.~S.}\ \bibnamefont {Bullock}},
  \bibinfo {author} {\bibfnamefont {S.}~\bibnamefont {Garrison-Kimmel}},
  \bibinfo {author} {\bibfnamefont {M.}~\bibnamefont {Boylan-Kolchin}},
  \bibinfo {author} {\bibfnamefont {M.~S.}\ \bibnamefont {Pawlowski}}, \ and\
  \bibinfo {author} {\bibfnamefont {A.~S.}\ \bibnamefont {Graus}},\ }\href@noop
  {} {\  (\bibinfo {year} {2018})},\ \Eprint {http://arxiv.org/abs/1811.12413}
  {arXiv:1811.12413 [astro-ph.GA]} \BibitemShut {NoStop}%
\bibitem [{\citenamefont {{Simon}}(2018)}]{2018ApJ...863...89S}%
  \BibitemOpen
  \bibfield  {author} {\bibinfo {author} {\bibfnamefont {J.~D.}\ \bibnamefont
  {{Simon}}},\ }\href {\doibase 10.3847/1538-4357/aacdfb} {\bibfield  {journal}
  {\bibinfo  {journal} {Astrophys. J.}\ }\textbf {\bibinfo {volume} {863}},\
  \bibinfo {eid} {89} (\bibinfo {year} {2018})},\ \Eprint
  {http://arxiv.org/abs/1804.10230} {arXiv:1804.10230} \BibitemShut {NoStop}%
\bibitem [{\citenamefont {Balberg}\ and\ \citenamefont
  {Shapiro}(2002)}]{Balberg:2001qg}%
  \BibitemOpen
  \bibfield  {author} {\bibinfo {author} {\bibfnamefont {S.}~\bibnamefont
  {Balberg}}\ and\ \bibinfo {author} {\bibfnamefont {S.~L.}\ \bibnamefont
  {Shapiro}},\ }\href {\doibase 10.1103/PhysRevLett.88.101301} {\bibfield
  {journal} {\bibinfo  {journal} {Phys. Rev. Lett.}\ }\textbf {\bibinfo
  {volume} {88}},\ \bibinfo {pages} {101301} (\bibinfo {year} {2002})},\
  \Eprint {http://arxiv.org/abs/astro-ph/0111176} {arXiv:astro-ph/0111176
  [astro-ph]} \BibitemShut {NoStop}%
\bibitem [{\citenamefont {Choquette}\ \emph {et~al.}(2018)\citenamefont
  {Choquette}, \citenamefont {Cline},\ and\ \citenamefont
  {Cornell}}]{Choquette:2018lvq}%
  \BibitemOpen
  \bibfield  {author} {\bibinfo {author} {\bibfnamefont {J.}~\bibnamefont
  {Choquette}}, \bibinfo {author} {\bibfnamefont {J.~M.}\ \bibnamefont
  {Cline}}, \ and\ \bibinfo {author} {\bibfnamefont {J.~M.}\ \bibnamefont
  {Cornell}},\ }\href@noop {} {\  (\bibinfo {year} {2018})},\ \Eprint
  {http://arxiv.org/abs/1812.05088} {arXiv:1812.05088 [astro-ph.CO]}
  \BibitemShut {NoStop}%
\bibitem [{\citenamefont {Chu}\ and\ \citenamefont
  {Garcia-Cely}(2018)}]{Chu:2018nki}%
  \BibitemOpen
  \bibfield  {author} {\bibinfo {author} {\bibfnamefont {X.}~\bibnamefont
  {Chu}}\ and\ \bibinfo {author} {\bibfnamefont {C.}~\bibnamefont
  {Garcia-Cely}},\ }\href {\doibase 10.1088/1475-7516/2018/07/013} {\bibfield
  {journal} {\bibinfo  {journal} {JCAP}\ }\textbf {\bibinfo {volume} {1807}},\
  \bibinfo {pages} {013} (\bibinfo {year} {2018})},\ \Eprint
  {http://arxiv.org/abs/1803.09762} {arXiv:1803.09762 [hep-ph]} \BibitemShut
  {NoStop}%
\bibitem [{\citenamefont {Vogelsberger}\ \emph {et~al.}(2018)\citenamefont
  {Vogelsberger}, \citenamefont {Zavala}, \citenamefont {Schutz},\ and\
  \citenamefont {Slatyer}}]{Vogelsberger:2018bok}%
  \BibitemOpen
  \bibfield  {author} {\bibinfo {author} {\bibfnamefont {M.}~\bibnamefont
  {Vogelsberger}}, \bibinfo {author} {\bibfnamefont {J.}~\bibnamefont
  {Zavala}}, \bibinfo {author} {\bibfnamefont {K.}~\bibnamefont {Schutz}}, \
  and\ \bibinfo {author} {\bibfnamefont {T.~R.}\ \bibnamefont {Slatyer}},\
  }\href@noop {} {\  (\bibinfo {year} {2018})},\ \Eprint
  {http://arxiv.org/abs/1805.03203} {arXiv:1805.03203 [astro-ph.GA]}
  \BibitemShut {NoStop}%
\bibitem [{\citenamefont {Essig}\ \emph {et~al.}(2018)\citenamefont {Essig},
  \citenamefont {Yu}, \citenamefont {Zhong},\ and\ \citenamefont
  {Mcdermott}}]{Essig:2018pzq}%
  \BibitemOpen
  \bibfield  {author} {\bibinfo {author} {\bibfnamefont {R.}~\bibnamefont
  {Essig}}, \bibinfo {author} {\bibfnamefont {H.-B.}\ \bibnamefont {Yu}},
  \bibinfo {author} {\bibfnamefont {Y.-M.}\ \bibnamefont {Zhong}}, \ and\
  \bibinfo {author} {\bibfnamefont {S.~D.}\ \bibnamefont {Mcdermott}},\
  }\href@noop {} {\  (\bibinfo {year} {2018})},\ \Eprint
  {http://arxiv.org/abs/1809.01144} {arXiv:1809.01144 [hep-ph]} \BibitemShut
  {NoStop}%
\end{thebibliography}%

\end{document}